\documentclass[reprint,preprintnumbers,amsmath,amssymb,superscriptaddress,nofootinbib,preprintnumbers,floatfix,aps,prc]{revtex4-2}

\usepackage{graphicx}% Include figure files
\usepackage{dcolumn}% Align table columns on decimal point

\usepackage{bigdelim}
\usepackage{amsmath}
\usepackage{amssymb}
\usepackage{lineno}
\usepackage{isotope}
\usepackage[unicode]{hyperref}
\usepackage{multirow}
\usepackage{xspace}

\usepackage[position=bottom, caption=false]{subfig}

\newcommand{\ve}[1]{\ensuremath{\mathbf{#1}}}
\def\p{\mbox{\boldmath $p$}}
\def\q{\mbox{\boldmath $q$}}
\def\P{\mbox{{\bf p}$^\prime$}}

\newcommand{\etal}{{\it et al.}\xspace}

\newcommand{\phantomsubfloat}[1]{
    {
        \captionsetup[subfigure]{labelformat=empty}
        \subfloat[][]{#1}
    }%
}

\hypersetup{
    colorlinks=true, %set true if you want colored links
    linkcolor=blue,  %choose some color if you want links to stand out
   citecolor=red
}

\begin{document}

\title{Determination of the argon spectral function from \texorpdfstring{$(e,e^\prime p)$}{(e,e'p)} data}
%measured in Jefferson Lab Hall A}

\author{L. Jiang} \affiliation{Center for Neutrino Physics, Virginia Tech, Blacksburg, Virginia 24061, USA}
\author{A.~M.~Ankowski} \affiliation{SLAC National Accelerator Laboratory, Stanford University, Menlo Park, California 94025, USA}
\author{D.~Abrams} \affiliation{Department of Physics, University of Virginia, Charlottesville, Virginia 22904, USA}
\author{L.~Gu} \affiliation{Center for Neutrino Physics, Virginia Tech, Blacksburg, Virginia 24061, USA}
\author{B.~Aljawrneh} \affiliation{North Carolina Agricultural and Technical State University, Greensboro, North Carolina 27401, USA}
\author{S.~Alsalmi} \affiliation{King Saud University, Riyadh 11451, Kingdom of Saudi Arabia}
\author{J.~Bane} \affiliation{The University of Tennessee, Knoxville, Tennessee 37996, USA}
\author{A.~Batz} \affiliation{The College of William and Mary, Williamsburg, Virginia 23187, USA}
\author{S.~Barcus} \affiliation{The College of William and Mary, Williamsburg, Virginia 23187, USA}
\author{M.~Barroso} \affiliation{Georgia Institute of Technology, Georgia 30332, USA}
\author{V.~Bellini} \affiliation{INFN, Sezione di Catania, Catania, 95123, Italy}
\author{O.~Benhar} \affiliation{INFN and Dipartimento di Fisica, Sapienza Universit\`{a} di Roma, I-00185 Roma, Italy}
\author{J.~Bericic} \affiliation{Thomas Jefferson National Accelerator Facility, Newport News, Virginia 23606, USA}
\author{D.~Biswas} \affiliation{Hampton University, Hampton, Virginia 23669, USA}
\author{A.~Camsonne} \affiliation{Thomas Jefferson National Accelerator Facility, Newport News, Virginia 23606, USA}
\author{J.~Castellanos} \affiliation{Florida International University, Miami, Florida 33181, USA}
\author{J.-P.~Chen} \affiliation{Thomas Jefferson National Accelerator Facility, Newport News, Virginia 23606, USA}
\author{M.~E.~Christy} \affiliation{Hampton University, Hampton, Virginia 23669, USA}
\author{K.~Craycraft} \affiliation{The University of Tennessee, Knoxville, Tennessee 37996, USA}
\author{R.~Cruz-Torres} \affiliation{Massachusetts Institute of Technology, Cambridge, Massachusetts 02139, USA}
\author{H.~Dai} \affiliation{Center for Neutrino Physics, Virginia Tech, Blacksburg, Virginia 24061, USA}
\author{D.~Day} \affiliation{Department of Physics, University of Virginia, Charlottesville, Virginia 22904, USA}
\author{A.~Dirican} \affiliation{Department of Physics, University of Maryland, College Park, MD 20742}
\author{S.-C.~Dusa} \affiliation{Thomas Jefferson National Accelerator Facility, Newport News, Virginia 23606, USA}
\author{E.~Fuchey} \affiliation{University of Connecticut, Storrs, Connecticut 06269, USA}
\author{T.~Gautam} \affiliation{Hampton University, Hampton, Virginia 23669, USA}
\author{C.~Giusti} \affiliation{Dipartimento di Fisica, Universit\`{a} degli Studi di Pavia and INFN, Sezione di Pavia,  I-27100 Pavia, Italy}
\author{J.~Gomez}\thanks{deceased} \affiliation{Thomas Jefferson National Accelerator Facility, Newport News, Virginia 23606, USA}
\author{C.~Gu} \affiliation{Duke University, Durham, North Carolina 27708, USA}
\author{T.~J.~Hague} \affiliation{Kent State University, Kent, Ohio 44242, USA}
\author{J.-O.~Hansen} \affiliation{Thomas Jefferson National Accelerator Facility, Newport News, Virginia 23606, USA}
\author{F.~Hauenstein} \affiliation{Old Dominion University, Norfolk, Virginia 23529, USA}
\author{D.~W.~Higinbotham} \affiliation{Thomas Jefferson National Accelerator Facility, Newport News, Virginia 23606, USA}
\author{C.~Hyde} \affiliation{Old Dominion University, Norfolk, Virginia 23529, USA}
\author{Z.~Jerzyk} \affiliation{Department of Physics, St. Norbert College, De Pere, Wisconsin 54115, USA}
\author{C.~Keppel} \affiliation{Thomas Jefferson National Accelerator Facility, Newport News, Virginia 23606, USA}
\author{S.~Li} \affiliation{University of New Hampshire, Durham, New Hampshire 03824, USA}
\author{R.~Lindgren} \affiliation{Department of Physics, University of Virginia, Charlottesville, Virginia 22904, USA}
\author{H.~Liu} \affiliation{Columbia University, New York, New York 10027, USA}
\author{C.~Mariani}\email{mariani@vt.edu} \affiliation{Center for Neutrino Physics, Virginia Tech, Blacksburg, Virginia 24061, USA}
\author{R.~E.~McClellan} \affiliation{Thomas Jefferson National Accelerator Facility, Newport News, Virginia 23606, USA}
\author{D.~Meekins} \affiliation{Thomas Jefferson National Accelerator Facility, Newport News, Virginia 23606, USA}
\author{R.~Michaels} \affiliation{Thomas Jefferson National Accelerator Facility, Newport News, Virginia 23606, USA}
\author{M.~Mihovilovic} \affiliation{Jozef Stefan Institute, Ljubljana 1000, Slovenia}
\author{M.~Murphy} \affiliation{Center for Neutrino Physics, Virginia Tech, Blacksburg, Virginia 24061, USA}
\author{D.~Nguyen} \affiliation{Department of Physics, University of Virginia, Charlottesville, Virginia 22904, USA}
\author{M.~Nycz} \affiliation{Kent State University, Kent, Ohio 44242, USA}
\author{L.~Ou} \affiliation{Massachusetts Institute of Technology, Cambridge, Massachusetts 02139, USA}
\author{B.~Pandey} \affiliation{Hampton University, Hampton, Virginia 23669, USA}
\author{V.~Pandey} \altaffiliation{Present Address: Department of Physics, University of Florida, Gainesville, FL 32611, USA}
\affiliation{Center for Neutrino Physics, Virginia Tech, Blacksburg, Virginia 24061, USA}
\author{K.~Park} \affiliation{Thomas Jefferson National Accelerator Facility, Newport News, Virginia 23606, USA}
\author{G.~Perera} \affiliation{Department of Physics, University of Virginia, Charlottesville, Virginia 22904, USA}
\author{A.~J.~R.~Puckett} \affiliation{University of Connecticut, Storrs, Connecticut 06269, USA}
\author{S.~N.~Santiesteban} \affiliation{University of New Hampshire, Durham, New Hampshire 03824, USA}
\author{S.~\v{S}irca} \affiliation{Faculty of Mathematics and Physics, University of Ljubljana, Ljubljana 1000, Slovenia} \affiliation{Jozef Stefan Institute, Ljubljana 1000, Slovenia}
\author{T.~Su} \affiliation{Kent State University, Kent, Ohio 44242, USA}
\author{L.~Tang} \affiliation{Hampton University, Hampton, Virginia 23669, USA}
\author{Y.~Tian} \affiliation{Shandong University, Shandong, 250000, China}
\author{N.~Ton} \affiliation{Department of Physics, University of Virginia, Charlottesville, Virginia 22904, USA}
\author{B.~Wojtsekhowski} \affiliation{Thomas Jefferson National Accelerator Facility, Newport News, Virginia 23606, USA}
\author{S.~Wood} \affiliation{Thomas Jefferson National Accelerator Facility, Newport News, Virginia 23606, USA}
\author{Z.~Ye} \affiliation{Physics Division, Argonne National Laboratory, Argonne, Illinois 60439, USA}
\author{J.~Zhang} \affiliation{Department of Physics, University of Virginia, Charlottesville, Virginia 22904, USA}

\collaboration{The Jefferson Lab Hall A Collaboration}
%-----------------------------------------------------------------------------------------------------------------------------------------------------------------------%

\begin{abstract}

The E12-14-012 experiment, performed in Jefferson Lab Hall A, has measured the $(e, e'p)$ cross section in parallel kinematics using a natural argon target. Here, we report the full results of the analysis of the data set corresponding to  beam energy 2.222~GeV, and spanning the missing momentum and missing energy range $15 \lesssim p_m \lesssim 300$ MeV/c and $12 \lesssim E_m \lesssim 80$ MeV. The reduced cross section, determined as a function of $p_m$ and $E_m$ with $\approx$4\% accuracy, has been fitted using the results of Monte Carlo simulations involving a model spectral function and including the effects of final state interactions. The overall agreement between data and simulations turns out to be quite satisfactory ($\chi^2$/d.o.f.=1.9). The resulting spectral function will provide valuable new information, needed for the interpretation of neutrino interactions in liquid argon detectors.

\end{abstract}

\preprint{JLAB-PHY-22-3575}
\preprint{SLAC-PUB-17650}
%
%-----------------------------------------------------------------------------------------------------------------------------------------------------------------------%
%
\maketitle
\section{Introduction}

Nucleon knockout reactions have long been recognized as a very  powerful tool to investigate
the properties of protons and neutrons bound in atomic nuclei; for a concise review see Ref.~\cite{Benhar_NPN}. In the kinematical conditions in which 
factorisation of the nuclear transition amplitude is applicable, the knockout 
cross section provides direct access to the nucleon momentum and removal energy distribution,  
described by the target spectral function.

Over the years, it has become also apparent that, in addition to being valuable in its own right, a quantitative  
understanding of nuclear spectral functions is 
required to interpret the
data collected by experiments in which nuclear interactions are exploited
to study the underlying processes involving nucleons. 
A notable example are accelerator-based searches of neutrino oscillations, in which an accurate 
description of neutrino-nucleus scattering is needed to determine the largely unknown properties of the beam particles~\cite{Benhar:2015wva}. 
 
The potential of knockout reactions was first established by pioneering studies of the $(p,2p)$ reaction carried out in the 1950s, although the interpretation of the measured cross sections was hindered by the occurrence of strong interactions involving both the incoming and outgoing protons. This problem severely limits access to the properties
of deeply bound nucleons in medium and heavy nuclei, because the dominant contributions to the $(p,2p)$ reaction
originate from interactions taking place on the nuclear surface.

The $(e,e^\prime p)$ process offers a clear advantage, 
because the projectile particle interacts weakly with the target, and probes the whole target volume. 
Therefore, the measured $(e,e^\prime p)$ cross sections have the potential to provide information on 
the proton spectral function up to the large values of removal energies corresponding to deeply bound states.

The conceptual framework underlying the determination of the spectral function from data is based on the 
assumptions that: (i) the beam-target interaction involves a single nucleon, and (ii) final state interactions (FSI) 
between the knocked out particle and the recoiling nucleus can be described in terms
of an optical potential within the Distorted Wave Impulse
Approximations (DWIA). Under these premises, the measured
 $(e,e^\prime p)$  cross section can be written in factorised form, see Eq.(1) below, 
and provides access to the nuclear spectral function. 

The validity of the above picture has been extensively discussed in the literature~\cite{Frullani:1984nn}.
The results of Ref.~\cite{MEC} indicate that the contribution
of processes involving two-nucleon Meson-Exchange Currents (MEC),
which are known to primarily affect the transverse components
of the nuclear response, become negligibly small in parallel kinematics.

Parallel kinematics has been also shown to preserve
factorisation of the $(e,e^\prime p)$ cross section in the presence of
sizable FSI effects~\cite{Boffi:1978sd}.

Early measurements of the  $(e,e^\prime p)$ reaction were aimed at testing the accuracy to which the 
data could be explained by the nuclear shell model, according to which the missing energy spectrum
consists of a collection of spectroscopic lines, corresponding to proton knockout from the
single-nucleon states belonging to Fermi sea of the target nucleus. Dynamical effects beyond the mean field approximation underlying the shell model 
are primarily associated with nucleon-nucleon correlations. The occurrence of correlations leads to a broadening of the spectroscopic lines and   
 to a significant quenching of the corresponding strengths\textemdash referred to as spectroscopic factors\textemdash which provide a measure of the occupation probability of shell model states.

The pattern described above has been clearly observed in $p$-shell nuclei, such as carbon and oxygen~\cite{Mougey:1976sc,Bernheim:1981si,VanDerSteenhoven:1988qa,Leuschner:1994zz}. The achieved experimental energy resolution allowed to unambiguously identify both position and width of the spectroscopic lines, and obtain the associated spectroscopic factors.  The spin-orbit splitting between the $p_{1/2}$ and 
$p_{3/2}$ levels was also determined with remarkable accuracy. 

The carbon spectral function has been also studied performing a 
measurement of the $(e,e^\prime p)$ cross section in the region of large missing momentum {\it and} large missing energy, where correlation effects play a critical role~\cite{daniela:1} and~\cite{daniela:2}. The integrated strength obtained from the experimental analysis, reported in Ref.~\cite{daniela:1}, turns out to be in remarkably good agreement with the results of theoretical calculations, based on realistic microscopic models of nuclear dynamics. Moreover, the amount of correlation strength is consistent with 
the quenching of spectroscopic factors resulting from the analysis of \isotope[12][]{C}$(e,e^\prime p)$ data at 
low missing energy and missing momentum.  

Systematic studies of the $(e,e^\prime p)$ reaction in heavier nuclei, ranging from \isotope[40][]{Ca} to \isotope[208][]{Pb},
have been also performed, mainly at Saclay and NIKHEF-K~\cite{Frullani:1984nn,Quint,Kramer}. However, owing to the more complex level scheme and 
to the large widths of the spectroscopic lines corresponding to deeply bound states, the individual contributions to the missing energy spectra significantly 
overlap with one another, and their identification requires a careful analysis, based on a detailed theoretical model 
of the spectral function.

\begin{table*}[tb]
\caption{\label{tab:kinematic}Kinematics settings used to collect the data analyzed here.}
\begin{ruledtabular}
\begin{tabular}{ c c c c c c c c c c c }
    & $E_{e}^\prime$      	& $\theta_e$      & $Q^2$             & $|\P|$         & $T_{p^\prime}$     & $\theta_{p^\prime}$     & $|{\bf q}|$     & $p_m$ & $E_m$ \\
    & (GeV)             		& (deg)         & (GeV$^2/c^2$)     & (MeV/$c$)     & (MeV) & (deg)                 &(MeV/$c$)     & (MeV/$c$) & (MeV)\\
    \colrule
    kin1 & 1.777             & 21.5         & 0.549             & 915         & 372              & $-50.0$              & 865          &   50 & 73  \\
    kin2 & 1.716             & 20.0         & 0.460             & 1030         & 455              & $-44.0$              & 846          &   184 & 50  \\
    kin3 & 1.799             & 17.5         & 0.370             & 915         & 372              & $-47.0$              & 741          &   174 & 50  \\
    kin4 & 1.799             & 15.5      & 0.291         & 915         & 372              &  $-44.5$             & 685          &   230            & 50  \\
    kin5 & 1.716             & 15.5        & 0.277         & 1030         & 455              &  $-39.0$             & 730          &   300            & 50  \\
\end{tabular}
\end{ruledtabular}
\end{table*} 
%%%%%%%%%%%%%%%%%%%%%%%%%%%%%%%%%%%%%%%%%%%%%%%%%%%%%%%%%%%%%%%%%%%%%%%%%%%%%%%%%%%%%%%%%

The primary goal of experiment E12-14-012~\cite{Benhar:2014nca,Dai:2018xhi,Dai:2018gch,Murphy:2019wed} is obtaining an accurate representation of the proton spectral functions of \isotope[40][18]{Ar} and \isotope[48][22]{Ti} from the corresponding $(e,e'p)$ cross sections, measured in Jefferson Lab Hall A~\cite{Gu:2020rcp}. The results of this study will provide the input needed to reconstruct the momentum and removal energy of the target nucleons, 
thus allowing for a reliable reconstruction of neutrino interactions in liquid-argon detectors.
 
The spectral function employed for the analysis described in this article, while being based on an admittedly simplified model, 
proved to be suitable to reproduce the argon data, and allows a smooth extrapolation to the region of low momenta, 
less than $\sim$15~MeV/c, not covered by our experiment. In addition, the results of the analysis appear to be consistent with the 
information obtained from previous studies of the $(p,2p)$ and $(d,\isotope[3]{He})$ reactions. 

A more detailed description of the spectral  function\textemdash including a precise determination of the 
energy and momentum distributions of the individual shell model states and the corresponding spectroscopic factors\textemdash 
will require the use of a more advanced theoretical model. Of great importance, in this context, will be the availability
of the results of {\it ab initio} calculations, discussed in Ref.~\cite{Barbieri:2019ual}. 
A comparison between the data collected using the \isotope[40][18]{Ar} and \isotope[48][22]{Ti} targets
will also allow to pin down effects arising from the isospin dependence of nuclear interactions.
The nuclear physics issues beyond the goal of experiment E12-14-012 will be the subject of forthcoming analyses.

The rest of this  article is organised as follows. The kinematic setup, the detectors and their resolutions, and our definitions of signal and backgrounds are briefly discussed in Sec.~\ref{sec:ExperimentalSetup}. In Sec.~\ref{sec:data_analysis}, we describe the procedure 
employed to obtain the reduced cross section, introduce the definitions of missing energy and missing momentum\textemdash which are the fundamental variables in our analysis\textemdash and describe the spectral function employed to model the data. Sec.~\ref{sec:Uncertainties} is devoted to the uncertainties associated with our analysis, while in Sec.~\ref{sec:comparison} we compare our results with the Monte Carlo predictions. In Sec.~\ref{sec:fit}, we describe the fitting procedure and results and in Sec.~\ref{sec:comparisonToPast} we compare our and previous experimental results. In Sec.~\ref{sec:Summary} we present a summary and our conclusions.

\section{Experimental Setup}\label{sec:ExperimentalSetup} 
Experiment E12-14-012 was approved by the Jefferson Lab PAC in 2014, and took data in Hall A in the Spring 2017. The measurements included the inclusive,  $(e,e^\prime)$~\cite{Dai:2018xhi,Dai:2018gch,Murphy:2019wed}, and exclusive $(e,e^\prime p)~$\cite{Gu:2020rcp} electron scattering cross sections on several targets, including a natural gas argon target~\cite{Gu:2020rcp}. 

The experiment was performed using an electron beam of energy 2.222~GeV provided by the Jefferson Lab Continuous Electron Beam Accelerator Facility (CEBAF), with an average beam current of 15~$\mu$A. The scattered electrons and protons were detected in coincidence in the Hall A high-resolution spectrometers (HRSs), using both the electron and proton spectrometers. Each spectrometer is equipped with vertical drift chambers (VDCs) providing tracking information~\cite{Fissum:2001st}, scintillator planes (two) for timing measurements and triggering, double-layered lead-glass calorimeter and a gas \v{C}erenkov counter used for particle identification~\cite{Alcorn:2004sb}. The electron arm is also equipped with pion rejectors~\cite{Alcorn:2004sb} while the proton arm contains a pre-shower and shower detectors~\cite{Alcorn:2004sb}. The experimental configurations of the HRSs in angle and momentum  are summarized in Table~\ref{tab:kinematic}. The beam current and position were monitored by resonant radio-frequency cavities (BCMs~\cite{Alcorn:2004sb}) and cavities with four antennae (BPMs~\cite{Alcorn:2004sb}), respectively. We used harp scanners (thin wire which moves  through the beam) to measure the beam size. The beam was spread on a $2\times2$ mm$^2$ area using a raster to avoid target overheating.

In addition to Ar and Ti, the experiment used C and Al targets to evaluate backgrounds and estimate the alignment and response of the spectrometer optics. We deployed two identical foils of the Al-7075 alloy (with thickness of $0.889\pm0.002$~g/cm$^2$) positioned to match the entrance and exit windows of the argon gas target cell.
The analysis discussed in this article used data collected at 5 different kinematics, summarized in Table~\ref{tab:kinematic}. In Table~\ref{tab:kinematic}, $E_{e}^\prime$ and $\theta_e$ denote the outgoing electron energy and the electron scattering angle,   $\p^\prime$,  $T_{p^\prime}$,  and $\theta_p$  the momentum,  kinetic energy, and angle  of the outgoing proton,  $p_m$ and $E_m$ the missing momentum and energy, $\q$ the momentum transfer and $Q^2$ the four-momentum transfer squared. All of our data were taken in parallel kinematics, with $\q$ parallel to $\P$. 
\par The interaction vertex, momentum and reconstructed direction (in-plane and out-of-plane angles) of the scattered electron and proton, were determined using the VDCs' tracking information. An optical matrix was built using carbon multi-foil data, and used to transform quantities between the focal plane and target~\cite{Dai:2018gch,Dai:2018xhi,Murphy:2019wed,Gu:2020rcp}. We have included variations of the optics and magnetic field in both HRSs in the analysis as systematic uncertainties.

The triggers were formed using the scintillator planes on both the electron and proton spectrometers, along with signals from the gas \v{C}erenkov (GC) detector, the pion rejector (PR), the pre-shower and the shower detector (PS). A detailed description of the trigger can be found in Ref.~\cite{Gu:2020rcp}. We required that the events originated within the central $\pm$10~cm of the gas target cell in order to avoid events coming from electron interactions in the target entry or exit windows. We verified that this contamination was negligible using also data from the tritium experiment at Jefferson Lab, which uses the same Al cell~\cite{Santiesteban:2018qwi}.
We calibrated the spectrometer optics using sieve slit measurements. The angle and positions of the spectrometers were surveyed at each kinematic settings. 
\par We studied in details the efficiencies of the elements in the detector stack by comparing rates in various combinations of secondary triggers as in Ref.~\cite{Dai:2018gch,Dai:2018xhi,Murphy:2019wed,Gu:2020rcp}.
We found an overall efficiency between 39.6\% and 48.9\% across all kinematic regions; for details, see Ref.~\cite{Gu:2020rcp}.

 The live-time of the electronics was estimated using scalers' rates. The acceptance cuts efficiencies were computed using simulated events~\cite{Arrington:1998ps}. The data was corrected for both nuclear transparency~\cite{Frankfurt:2000ty,Arrington:1998ps} and final state interactions (FSI)~\cite{Gu:2020rcp}. %
 
The energy resolution (typically 6$-$7~MeV) is affected by the uncertainty in the vertex resolution, Coulomb corrections, energy loss due to use of an extended argon gas target, and boiling effect from using a closed cell target. It is lower than what has been achieved by previous Hall A experiments but we would like to point out that, in the presence of large overlaps between proton orbital shells, the identification of different shell contributions to the missing energy spectrum relies mainly on the availability of a reliable model of the spectral function~\cite{Kramer:1990}.

%%%%%%%%%%%%%%%%%%%%%%%%%%%%%%%%%%%%%%%%%%%%%%%%%%%%%%%%%%%%%%%%%%%%%%%%%%%%%%%%%%%%%%%%%
%
\section{Data Analysis}\label{sec:data_analysis}
\subsection{The \texorpdfstring{$(e,e^\prime p)$}{(e,e'p)} reduced cross section}

The factorisation {\it }ansatz\textemdash whose validity is discussed in detail in Refs.~\cite{Boffi:1978sd,Boffi:1993gs,bof96}\textemdash provides the basis for the extraction of the target spectral 
function from $(e,e^\prime p)$ data. Within the factorised DWIA, the measured cross section can 
be written in the form 
\begin{equation}
\frac{d^6 \sigma }{ d\omega d\Omega_{ e^\prime} d T_{p^\prime} d\Omega_{p^\prime}} = K \sigma_{ep} P^D(\P,-\p_m,E_m).
\label{eq:dsf}
\end{equation}
where the solid angles $\Omega_{ e^\prime}$ and $\Omega_{ p^\prime}$ specify the directions of the 
outgoing particles, while $\omega$ and $T_{p^\prime}$ denote the energy transfer and
the kinetic energy of the knocked out proton, respectively. In the right-hand side of the above equation, 
$K = |\P| E_{p^\prime}$, with $E_{p^\prime} = T_{p^\prime} + M$, $M$ being the proton mass, and
$\sigma_{ep}$ is the differential cross section describing electron scattering off a bound moving proton.

Within this scheme the nucleon 
spectral function can be obtained by removing from 
$P^D(\P,-\p_m,E_m)$ the distortions arising from FSI effects. The missing momentum and missing energy are defined as
\begin{equation}
\label{pmiss}
        \p_m = \q-\P= \p_R,
\end{equation}
and
\begin{equation}
\label{Emiss}
        E_m=\omega-T_{p^\prime}-T_R,
\end{equation}
with $\p_R$ and $T_R$ being the momentum and kinetic energy of the recoiling residual nucleus.

The six-fold differential cross section as a function of $p_{m}$ and $E_{m}$ was extracted from the data using the $(e,e^\prime p)$ event yield $Y$ for each $p_{m}$ and $E_{m}$ bin, 
\begin{equation}
\frac{d^6 \sigma }{ d\omega d\Omega_{ e^\prime} d T_{p^\prime} d\Omega_{p^\prime}}  = \frac{{Y}(p_{m},E_{m})}{B \times lt \times \rho \times BH \times V_B \times C_\text{rad}}.
\label{eq:yield}
\end{equation}
Here, $B$ is the total accumulated beam charge; $lt$ is the live-time of the detector (fraction of time that the detector was able to collect and write data to disk); $\rho$ is the target density (for argon, corrected for the nominal density of gas in the target cell); $BH$ is the local density change due to the beam heating the gas cell times the gas expansion due to boiling effects; $V_B$ is the effect of the acceptance and kinematical cuts; finally, $C_\text{rad}$ is the effect of the radiative corrections and bin center migration.

The reduced cross section, identified with the distorted spectral function of Eq.~(\ref{eq:dsf}), has been obtained from the measured cross section as a function of $p_m=|\p_m|$
and $E_m$, by dividing out the kinematic factor $K$ and the electron-proton cross section $\sigma_{ep}$.  The off-shell extrapolation of de Forest~\cite{, Dieperink:1976wy,DeForest:1983ahx} has been adopted to describe the off-shell proton  cross section.

The SIMC spectrometer package~\cite{SIMC} was used to simulate $(e,e^\prime p)$ events corresponding to our kinematic settings, including geometric details of the target cell, radiation correction, and Coulomb effects. SIMC also provides the $V_B$ and $C_\text{rad}$ appearing in Eq.~\eqref{eq:yield}.  We used an approximate spectral function (SF)~\cite{Gu:2020rcp} as input to SIMC to simulate events. Unlike the test SF described in Sec.~\ref{sec:SF}, this SF model does not account for correlations between nucleons and assumes that the shell-model states are fully occupied, as prescribed by the independent-particle shell model. 

Monte Carlo (MC) events are generated over a broad phase-space, and propagate through a detailed model of the electron and proton spectrometers, accounting for acceptances and resolution effects. The events are weighted by the $\sigma_{cc1}$ cross section of de~Forest~\cite{DeForest:1983ahx} and the SF. The final weighted events are then background subtracted. We estimated the background performing analysis for each bin of $E_m$ (1~MeV) and $p_m$ (1~MeV/c). We use events selected in anti-coincidence between the electron and proton arms. This region corresponds to 100 times the nominal coincidence time window width that was set to $\approx$2~ns~\cite{Gu:2020rcp}. The events are then re-scaled based on the width of the coincidence peak. The background-event distributions were then generated and subtracted bin by bin from the $E_m$ and $p_m$ distributions.

%%%%%%%%%%%%%%%%%%%%%%%%%%%%%%%%%%%%%%%%%%%%%%%%%%%%%%%%%%%%%%%%%%%%%%%%%%%%%%%

\subsection{Test spectral function}\label{sec:SF}
In general, the spectral function could be decomposed into mean-field and correlation components,
\begin{equation}\label{eq:SF}
P(p_m,E_m) = P_\text{MF}(p_m,E_m)+P_\text{corr}(p_m,E_m).
\end{equation}
In constructing the test spectral function, we express its mean field part as a sum of the contributions of the states expected to be occupied in the independent-particle shell model
\begin{equation}\label{eq:MFSF}
P_\text{MF}(p_m,E_m) = \sum_{\alpha} S_{\alpha} |\phi_\alpha(p_m)|^2 f_\alpha(E_m) . 
\end{equation}
Here, $S_{\alpha}$ denotes the spectroscopic factor of the state $\alpha$, whose unit-normalised  momentum-space wave function is $\phi_{\alpha}(p_m)$, while $f_\alpha(E_m)$ is the corresponding missing energy distribution.

\begin{figure}[tb]
    \centering
    \includegraphics[width=0.8\columnwidth]{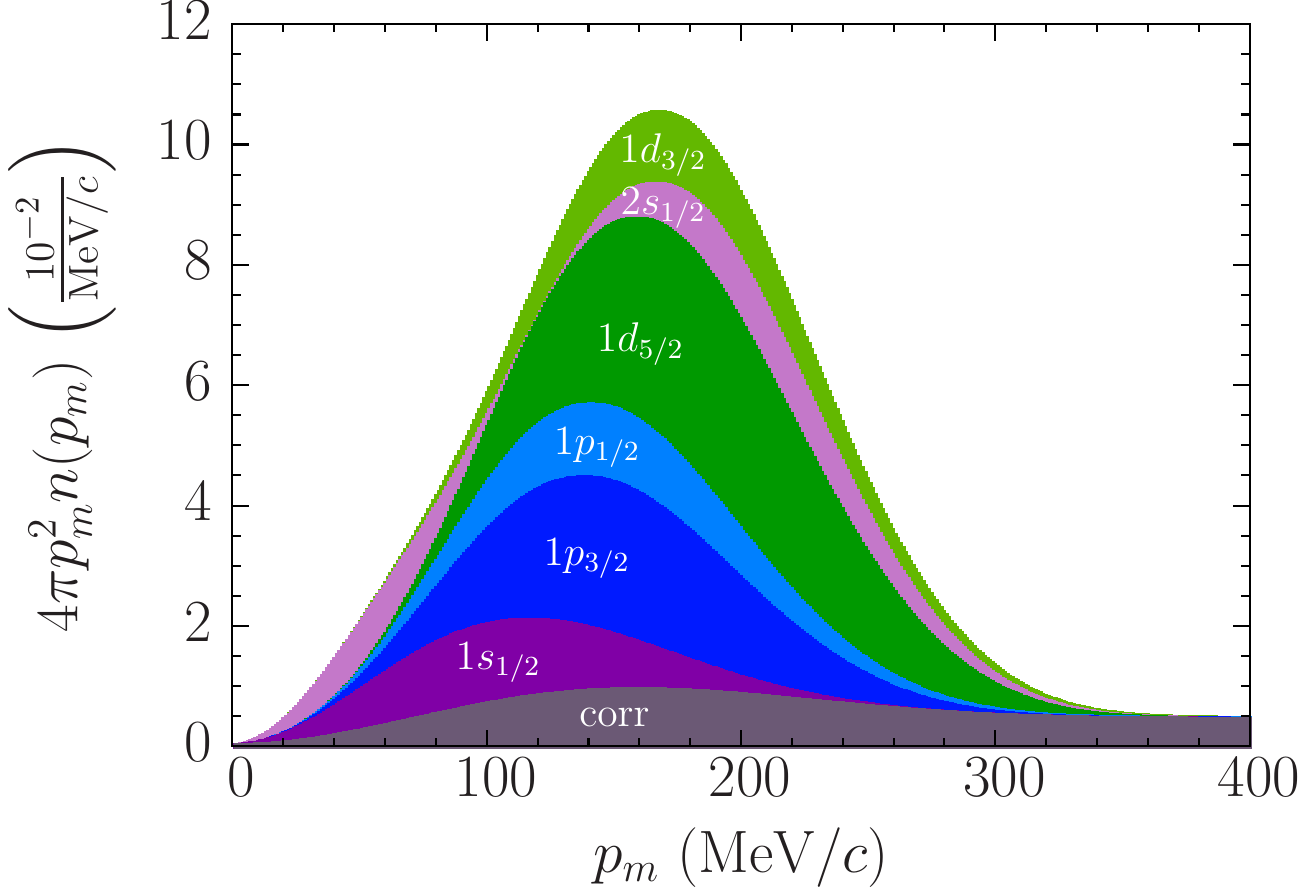}
    \caption{Missing momentum distribution of protons in argon in the test spectral function, presented with the geometric factor of $4\pi p_m^2$.}
    \label{fig:missing_momentum_distribution}
\end{figure}

In order to approximately account for the depletion of the shell-model states arising from short-range nucleon-nucleon correlations, the values of the spectroscopic factors are set to $S_\alpha=0.8N_\alpha$, $N_\alpha$ being the occupation number of the state $\alpha$ in the independent-particle shell model. For fully occupied shells, $N_\alpha=2j+1$, with $j$ being the corresponding total angular momentum.

We employ the wave functions of Ref.~\cite{Niksic:2014dra}. The resulting momentum distribution is presented in Fig.~\ref{fig:missing_momentum_distribution}, which shows that states of different orbital momentum tend to peak at different values of missing momentum.

\begin{table}[tb]
    \centering
    \caption{\label{tab:ar_energy_levels}Parametrization of the test spectral function of protons in argon. For each shell-model state $\alpha$, we compare the occupation number in the independent particle shell-model $N_\alpha$ with the assumed spectroscopic factor $S_\alpha$. The peak of the missing energy distribution $E_{\alpha}$ of the width $\sigma_{\alpha}$ is also provided. For the correlated part, we give its total normalization and the threshold for two-nucleon knockout $E_\text{thr}$.}
    \begin{ruledtabular}
    \begin{tabular}{ c d d d d }
     $\alpha$  & \multicolumn{1}{c}{$N_\alpha$} & \multicolumn{1}{c}{$S_\alpha$} & \multicolumn{1}{c}{$E_\alpha$ (MeV)} & \multicolumn{1}{c}{$\sigma_\alpha$ (MeV)} \\
     \colrule
       $1d_{3/2}$ & 2 & 1.6 & 12.53 & 2 \\
       $2s_{1/2}$ & 2 & 1.6 & 12.93 & 2 \\
       $1d_{5/2}$ & 6 & 4.8 & 18.23 & 4 \\
       $1p_{1/2}$ & 2 & 1.6 & 28.0 & 6  \\
       $1p_{3/2}$ & 4 & 3.2 & 33.0 & 6 \\
       $1s_{1/2}$ & 2 & 1.6 & 52.0 & 10 \\
       corr. & \text{---} & 3.6 & 20.60 & \text{---} \\
    \end{tabular}
    \end{ruledtabular}
\end{table}

\begin{figure}
    \centering
    \includegraphics[width=0.8\columnwidth]{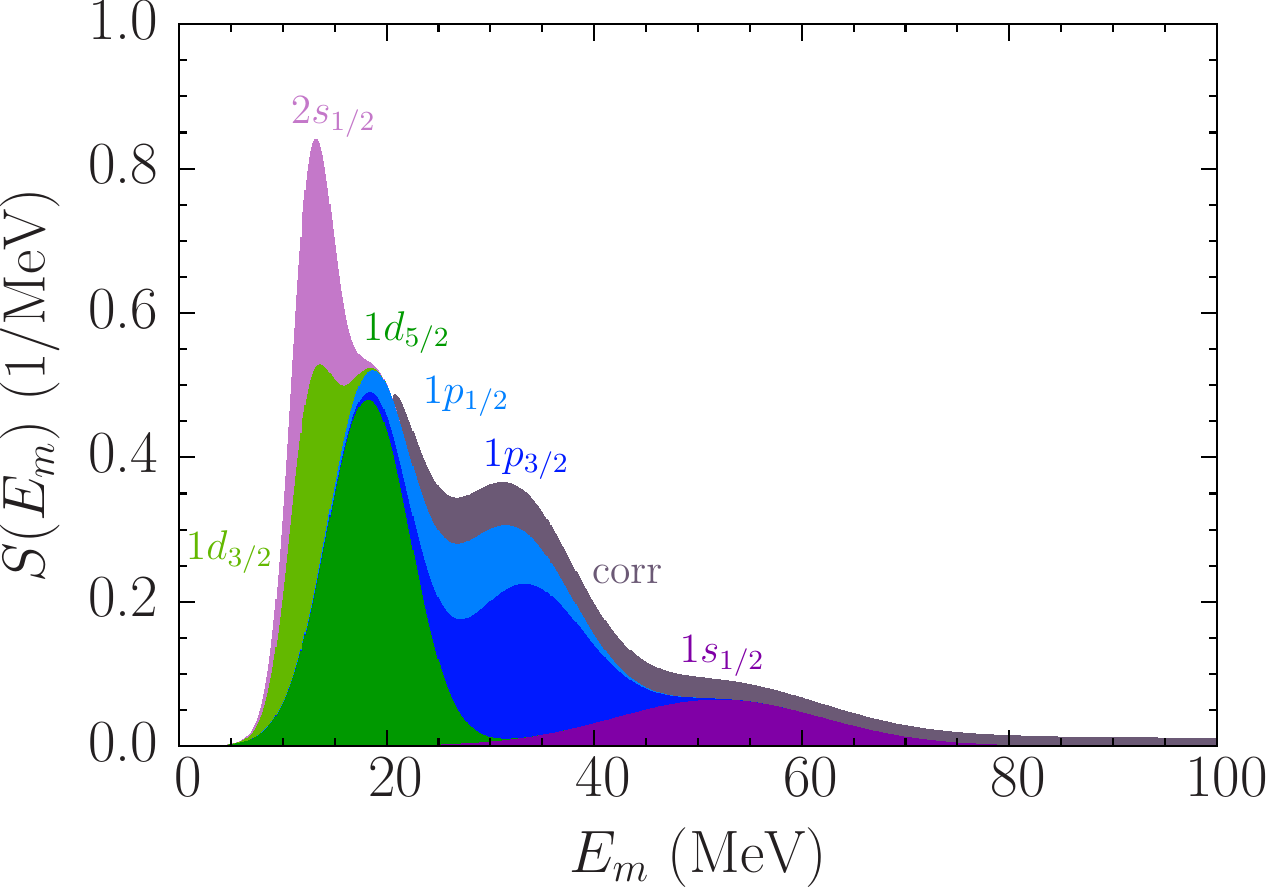}
    \caption{Missing energy distribution of protons in argon in the test spectral function.}
    \label{fig:missing_energy_distribution}
\end{figure}

The missing energy of the shell-model states is assumed to follow the Gaussian distribution,
\begin{equation}\label{eq:gauss}
f_\alpha(E_m) = \frac{1}{\sqrt{2\pi}\sigma_\alpha} \exp \left [-\left (\frac{E_m-E_\alpha}{\sqrt{2}\sigma_\alpha}\right )^2 \right ],
\end{equation}
peaked at the value $E_\alpha$ and with the width governed by $\sigma_\alpha$. All the parameters of the mean-field spectral function are provided in Table~\ref{tab:ar_energy_levels}. The resulting missing energy distribution is presented in Fig.~\ref{fig:missing_energy_distribution}.

To estimate the correlated spectral function we follow the approach of Ref.~\cite{CiofidegliAtti:1995qe}.
We express it as a convolution integral involving the momentum distributions of the relative and center-of-mass motion of a correlated proton-neutron ($pn$)
pair,
\begin{equation}\label{eq:corrSF}
\begin{split}
P_\text{corr}(p_m,E_m) &=\int d^3p_{A-2}\,\delta\left(E_m - E_\text{corr}-T_{A-1}\right)\\
&\times n^{pn}_\text{cm}(|\ve p_{A-2}|)n_\text{rel}^{pn}\left(\Big|\ve p_m+\frac{\ve p_{A-2}}2\Big|\right),
\end{split}
\end{equation}
where $E_\text{corr}$ is set to the value of the $pn$ knockout threshold, $E_\text{thr}=20.60$ MeV, estimated using the masses of the \isotope[38][17]{Cl} and \isotope[40][18]{Ar} nuclei~\cite{Wang:2017}, and $T_{A-1}$ is the energy of the relative motion of the correlated neutron and the $(A-2)$-nucleon system,
\[
T_{A-1}=\frac{A-2}{2M(A-1)}\left[\ve p_m+\frac{(A-1)\ve p_{A-2}}{A-2}\right]^2.
\]

\begin{figure}[bt]
    \centering
    \phantomsubfloat{\label{fig:SFa}}
    \phantomsubfloat{\label{fig:SFb}}
    \includegraphics[width=0.8\columnwidth]{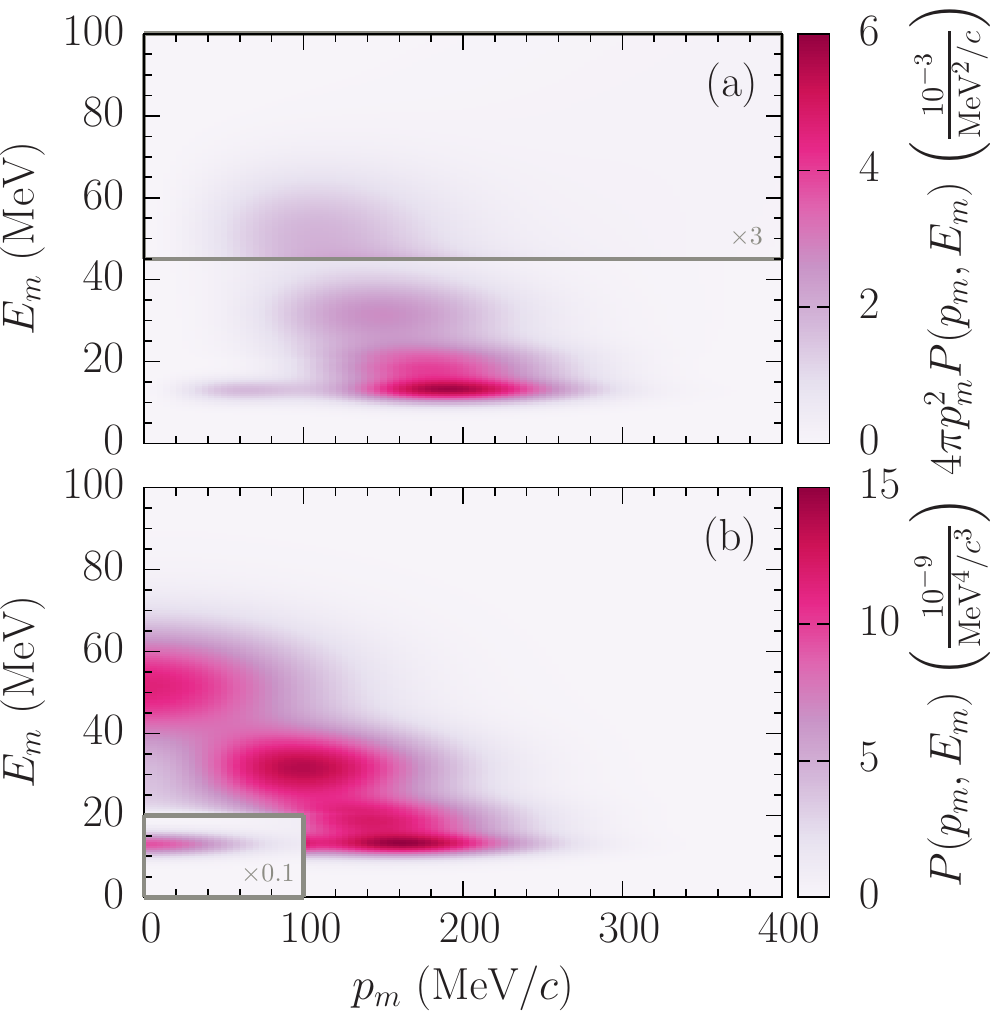}
    \caption{Test spectral function shown (a) with and (b) without the geometric factor of $4\pi p_m^2$. Note that multiplicative factors are used for clearer presentation of certain regions.}
\label{fig:SF}
\end{figure}

The center-of-mass momentum distribution of the $pn$ pair is expressed as
\begin{equation}
n^{pn}_\text{cm}(p)=Z\left(\frac{\alpha_\text{cm}}{\pi}\right)^{3/2}\exp(-\alpha_\text{cm} p^2),
\end{equation}
with $Z$ being the charge number. For the argon nucleus we employ $\alpha_\text{cm}=0.98$ fm$^2$, given in Ref.~\cite{CiofidegliAtti:1995qe} for \isotope[40][20]{Ca}.

The relative momentum distribution of the $pn$ pair is assumed to be the sum of two Gaussians,
\begin{equation}
n_\text{rel}^{pn}(p)=\frac{C^A}{4\pi}\left[A_1\exp(-\alpha_1 p^2)+A_2\exp(-\alpha_2p^2)\right],
\end{equation}
with the scaling factor $C^A=4.4$~\cite{CiofidegliAtti:1995qe}. We determine their parameter values to be $A_1=0.23444$ fm$^3$, $\alpha_1=3.2272$ fm$^2$, $A_2=0.006989$ fm$^3$, $\alpha_2=0.23308$ fm$^2$, by requiring the correlated momentum distribution to match the corresponding distribution  of \isotope[40][20]{Ca} reported in Ref.~\cite{CiofidegliAtti:1995qe}. 

By construction, the correlated part accounts for 20\% of the total strength of the test spectral function, see Table~\ref{tab:ar_energy_levels}. Of these $0.2Z=3.60$ protons, 1.99 (1.71) [1.35] come from the area defined by $p_m\leq400$ (300) [230] MeV and $E_m\leq100$ MeV. For comparison, the mean-field part predicts 14.40 (14.22) [12.30] protons in the same area.

Figure~\ref{fig:SF} displays the test spectral function as a function of missing momentum and missing energy. 
For comparison, we show both  $P(p_m, E_m)$ and $4\pi p_m^2P(p_m, E_m)$. In the region $E_m\geq45$ MeV of panel~\subref{fig:SFa}, the result is tripled to highlight the broad peak of the $1s_{1/2}$ shell of the spectral function multiplied by the phase-space factor. Conversely, a multiplicative factor 0.1 is applied in the region of $p_m\leq100$ MeV and $E_m\leq20$ MeV of panel~\subref{fig:SFb}, so that the narrow peak of the $2s_{1/2}$ shell does not dominate the landscape of the bare spectral function.

 \begin{figure}
    \centering
    \includegraphics[width=0.8\columnwidth]{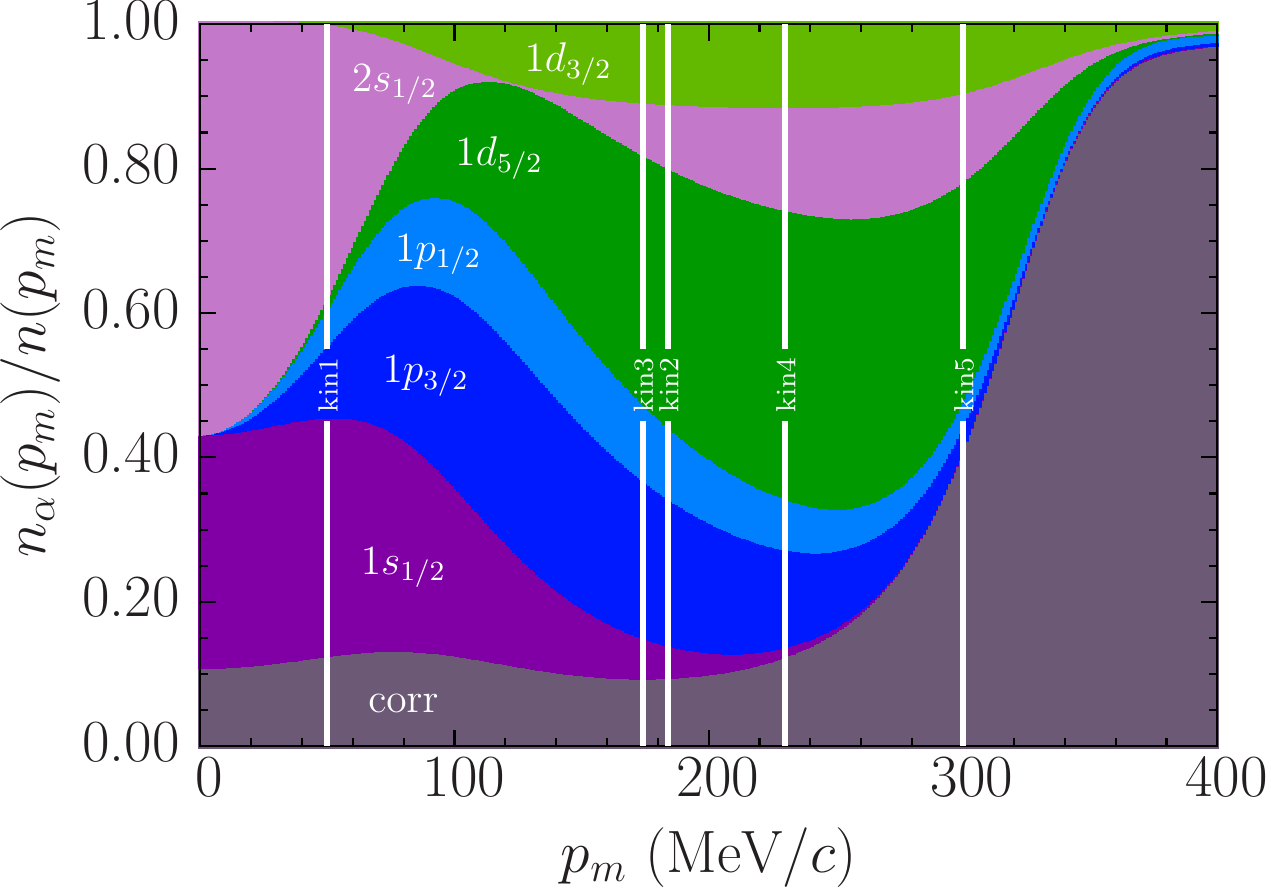}
    \caption{Fractional contributions to the missing momentum distribution at the kinematic settings of this experiment.}
    \label{fig:missing_momentum_contributions_for_setups}
\end{figure}

\begin{figure}[tb]
    \centering
    \includegraphics[width=0.8\columnwidth]{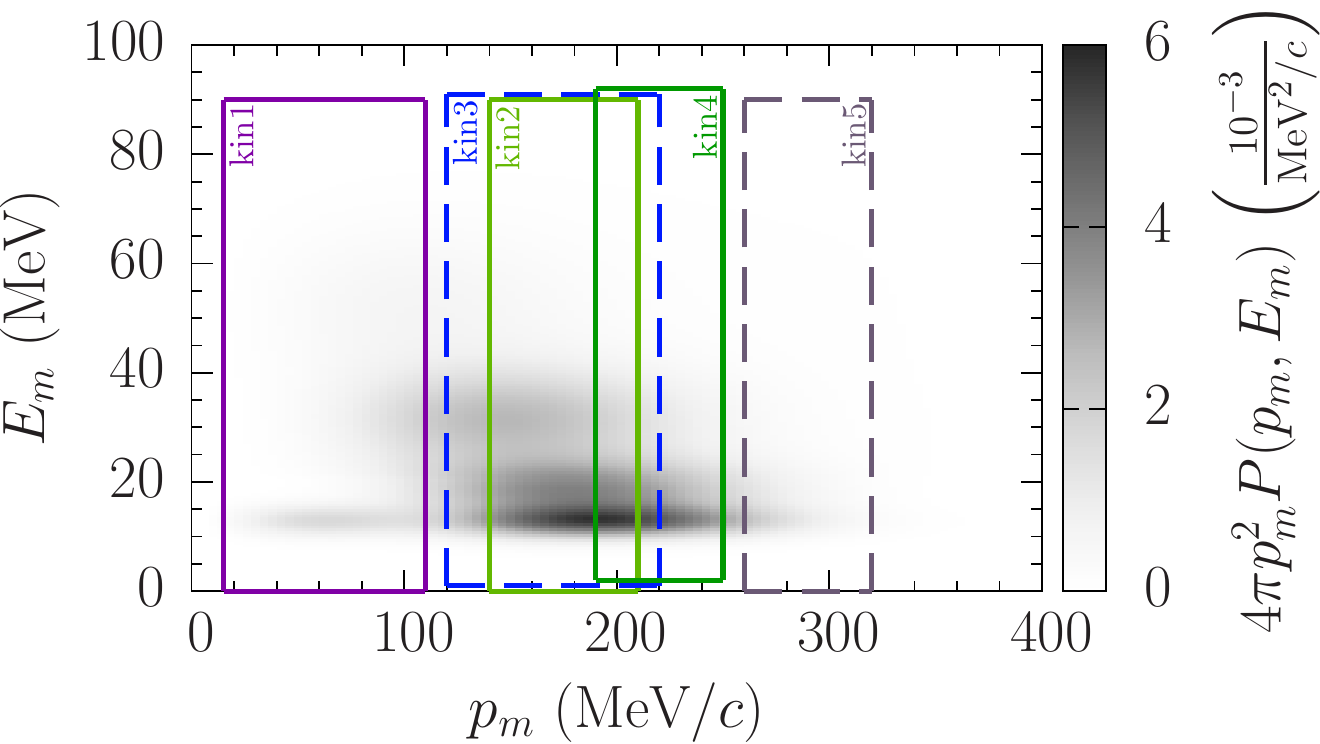}
    \caption{Kinematic coverage of this experiment overlaid on the test spectral function. The window representing kin3 (kin4) is shifted vertically by $+1$ ($+2$) MeV for clearer presentation. The coverage accounts for detector effects.}
\label{fig:Ar_kins}
\end{figure}

\begin{table*}[tb]
    \centering
    \caption{\label{tab:ar_S_contributions}Estimate of the spectroscopic strengths probed at each kinematics using the test spectral function. For clarity, only nonvanishing entries are shown. The total strength accessible in the experiment is 14.96.}
    \begin{ruledtabular}
    \begin{tabular}{ c r r c c c c c c c c}
& $|\ve p_m|$ range & $E_m$ range & $1d_{3/2}$ & $2s_{1/2}$ & $1d_{5/2}$ & $1p_{1/2}$ & $1p_{3/2}$ & $1s_{1/2}$ & \text{corr.} & \text{sum} \\
 & (MeV$/c$) & (MeV) & & & & & & & & \\
     \colrule
\multirow[c]{3}{*}{kin1} & \multirow[c]{3}{*}{15--110} & 0--30 & 0.11 & 0.48 & 0.26 & 0.18 & 0.18 & 0.01 & 0.11 & 1.34\\
& & 30--54 &  &  & & 0.11 & 0.40 & 0.44 & 0.18 & 1.13\\
& & 54--90 &  &  & & & & 0.33 & 0.05 & 0.38\\[3pt]
\multirow[c]{3}{*}{kin2} &  \multirow[c]{3}{*}{140--210} & 0--30 & 0.79 & 0.51 & 2.37 & 0.46 & 0.46 & 0.01 & 0.18 & 4.77\\
& & 30--54 &  &  & & 0.27 & 1.03 & 0.23 & 0.31 & 1.85\\
& & 54--90 &  &  & & & & 0.17 & 0.10 & 0.28\\[3pt]
\multirow[c]{3}{*}{kin3} &  \multirow[c]{3}{*}{120--220} & 0--30 & 1.04 & 0.62 & 3.11 & 0.64 & 0.64 & 0.01 & 0.26 & 6.33\\
& & 30--54 & & & & 0.38 & 1.44 & 0.37 & 0.44 & 2.64\\
& & 54--90 & & & & & & 0.28 & 0.14 & 0.42\\[3pt]
\multirow[c]{3}{*}{kin4} &  \multirow[c]{3}{*}{190--250} & 0--30 & 0.52 & 0.56 & 1.68 & 0.21 & 0.21 & & 0.12 & 3.30\\
& & 30--54 & & & & 0.12 & 0.47 & 0.05 & 0.21 & 0.86\\
& & 54--90 & & & & & & 0.04 & 0.09 & 0.13\\[3pt]
\multirow[c]{3}{*}{kin5} &  \multirow[c]{3}{*}{260--320} & 0--30 & 0.12 & 0.15 & 0.39 & 0.03 & 0.02 & & 0.05 & 0.76\\
& & 30--54 & & & & 0.02 & 0.05 & & 0.11 & 0.17\\
& & 54--90 & & & & & & & 0.09 & 0.09\\
    \end{tabular}
    \end{ruledtabular}
\end{table*}

The kinematic setup of this experiment~\cite{Benhar:2014nca} was designed to allow the separation of individual contributions to the spectral function, and for cross-checking the accuracy of the employed description of FSI. This feature is schematically illustrated in Fig.~\ref{fig:missing_momentum_contributions_for_setups}, which depicts the fractional contributions to the missing momentum distribution. For the state $\alpha$ of the mean-field component, such contribution can be expressed as
\begin{equation}
\frac{ n_\alpha(p_m) }{ n(p_m) } = \frac{ S_\alpha|\phi_\alpha(p_m)|^2} 
{\int_0^\infty P(p_m,E_m) \ dE_m } \ .
\end{equation}
Because this experiment covers a broad range of missing energies, extending from 0 to 90 MeV, this simple expression is a good approximation of the actual contributions of the shell-model states. One needs to bear in mind, however, that the correlated contribution obtained in this manner is overestimated at high missing momenta, where the strength for $E_m>90$ MeV plays non-negligible role.

The actual kinematic coverage in the $(p_m,E_m)$ space is presented in Fig.~\ref{fig:Ar_kins}. Table~\ref{tab:ar_S_contributions} details our estimates of the spectroscopic strengths for every kinematic setting calculated using the test spectral function.

%%%%%%%%%%%%%%%%%%%%%%%%%%%%%%%%%%%%%%%%%%%%%%%%%%%%%%%%%%%%%%%%%%%%%%%%%%%%%%%%

\section{Uncertainty Analysis}\label{sec:Uncertainties}
\begin{table*}[tb!]
\caption{\label{tab:syst}Contributions to systematical uncertainties for argon average over all the $E_m$ and $p_m$ bins for each kinematic. All numbers are in \%. kin4 and kin5 systematic is the sum in quadrature of the systematic uncertainties on the signal and the background.}
\begin{ruledtabular}
\begin{tabular}{@{}l c c c c c c @{}}
					&						            	& kin1     		& kin2		& kin3		& kin4		& kin5   	\\
\hline
{1.~Total statistical uncertainty} &			        	& 0.53   		& 0.57		& 0.64		& 0.54		&1.65	\\
{2.~Total systematic uncertainty}  &			        	& 3.14	   		& 3.24		& 3.32		& 10.23		& 9.01	\\
\phantom{1. }a.~Beam $x\&y$ offset &						& 0.63			& 0.85		& 0.69		& 0.91		& 1.68	\\
\phantom{1. }b.~Beam energy & 					    		& 0.10			& 0.10		& 0.10		& 0.10		& 0.10	\\
\phantom{1. }c.~Beam charge &					    		& 0.30			& 0.30		& 0.30		& 0.30		& 0.30  	\\
\phantom{1. }d.~HRS $x\&y$ offset &				    		& 0.83			& 1.17		& 0.78		& 1.44		& 1.71	\\
\phantom{1. }g.~Optics (q1, q2, q3) &						& 0.94			& 0.77		& 0.55		& 0.90		& 1.72	\\
\phantom{1. }h.~Acceptance cut $(\theta,\phi, z)$ & 		& 1.16			& 1.33		& 1.75		& 2.19		& 7.72	\\
\phantom{1. }i.~Target thickness/density/length & 			& 0.20			& 0.20		& 0.20		& 0.20		& 0.20	\\
\phantom{1. }j.~Calorimeter \& \v{C}erenkov cut & 			& 0.02			& 0.02		& 0.02		& 0.02		& 0.02	\\
\phantom{1. }k.~Radiative and Coulomb corr. & 				& 1.00			& 1.00		& 1.00		& 1.00		& 1.00	\\
\phantom{1. }l.~$\beta$ cut		 &				    		& 0.47			& 0.55		& 0.39		& 7.74		& 5.87	\\
\phantom{1. }m.~Boiling effect		&						& 0.70			& 0.70		& 0.70		& 0.70		& 0.70	\\
\phantom{1. }n.~Cross section model &						& 1.00			& 1.00		& 1.00		& 1.00		& 1.00	\\
\phantom{1. }o.~Trigger and coincidence time cut &			& 0.92			& 0.52		& 0.98		& 5.55		& 2.58	\\
\phantom{1. }p.~FSI				 &						    & 2.00			& 2.00		& 2.00		& 2.00		& 2.00	\\
\end{tabular}
\end{ruledtabular}
\end{table*}

The total systematic uncertainty in this analysis was estimated for each of the kinematics by summing in quadrature the individual uncertainties as listed in Table~\ref{tab:syst}. For kin4 and kin5, the systematic uncertainties were evaluated for the signal and background separately and then summed in quadrature. Kinematic and acceptance cuts are uncorrelated bin to bin and they do not depend on the theoretical input model. All the kinematic and acceptance cuts were varied by the resolution of the variable under consideration. The MC employed to evaluate those uncertainties did not contain FSI effects other than the transparency corrections. The MC simulation could depend a priori on the underlying theoretical model. However, we repeated the analysis of systematic uncertainties varying its ingredients, and did not observe any substantial variations of the obtained results. To determine the uncertainties related to the target position, we performed the simulation with the inputs for the beam's and spectrometer's $x$ and $y$ offsets varied within uncertainties, and we recomputed the optical transport matrix varying the three quadrupole magnetic fields, one at the time. Each of these runs was compared to the reference run, and the corresponding differences were summed in quadrature to give the total systematic uncertainty due to the Monte Carlo simulation as described in Ref.~\cite{Gu:2020rcp}.

The calorimeter and \v{C}e{\-}ren{\-}kov detector uncertainties were determined by changing the corresponding cut by a small amount and calculating the difference with respect to the nominal yield value. We computed the uncertainty due to the acceptance cuts on the angles in the same way. We included a fixed uncertainty for the beam charge and beam energy, as well as for the boiling effect, radiative and Coulomb corrections. The FSI uncertainty is estimated to be in the range  1--3\% depending on the kinematics~\cite{Gu:2020rcp}. The systematic uncertainties related to the trigger efficiency were determined across multiple runs, and by applying different acceptance cuts. A fixed uncertainty was assigned to take care of those variations. The time-coincidence cut efficiency was evaluated by changing the cut by $\pm \sigma$.

SIMC generates events including the effects from radiative processes: vacuum polarization, vertex corrections, and internal bremsstrahlung. In SIMC, radiative correction are determined following Dasu~\cite{Dasu}, who uses Whitlow's approach~\cite{Mo:1968cg,Whitlow:1990}. We considered a fixed 1\% uncertainty due to the theoretical model for the radiative corrections over the full kinematic range as in our previous work~\cite{Gu:2020rcp}. A target thickness uncertainty is also considered together with an uncertainty due to the boiling effect correction~\cite{Santiesteban:2018qwi}.

%%%%%%%%%%%%%%%%%%%%%%%%%%%%%%%%%%%%%%%%%%%%%%%%%%%%%%%%%%%%%%%%%%%%%%%%%%%%

\subsection{Final-state interactions}\label{sec:FSI}

In the distorted-wave impulse approximation (DWIA), FSI between the outgoing proton and the spectator nucleons are described by a complex, energy-dependent, phenomenological optical potential. We evaluate FSI by performing relativistic DWIA calculations following Ref.~\cite{Meucci:2001qc}.

To correct our MC events we have used a ``democratic''(DEM) relativistic optical potential~\cite{Cooper:2009zza}, obtained from a global fit to over 200 sets of elastic proton-nucleus scattering data, comprising a broad range of targets from helium to lead, at kinetic energies up to 1040~MeV.

The quenching of the calculated cross section resulting from FSI was computed using the DWIA/PWIA ratio, that is, the ratio of the integral over $p_m$ of the DWIA and PWIA (Plane Wave Impulse Approximation) reduced cross sections. Both the shift and the DWIA/PWIA ratios are computed separately for the positive and negative $p_m$ regions, 
corresponding to $|\q|<|\p^\prime|$ and $|\q|>|\p^\prime|$, respectively.
The theoretical uncertainty of the shift and the reduction produced by FSI has been evaluated and is described in details in Ref.~\cite{Gu:2020rcp}.

In our analysis, for each kinematic setting and for each event, the FSI corrections have been applied in both the missing energy and missing momentum distributions. 

For each event, we determine the orbital involved using the reconstructed energy and momentum of both electron and proton, and apply the corresponding FSI correction. For overlapping orbitals,
we use a simple prescription to determine the most probable orbital from which the electron was emitted, as described in Ref.~\cite{Gu:2020rcp}.

\begin{figure}[tb]
\subfloat[Data]{\includegraphics[width=1.0\columnwidth]{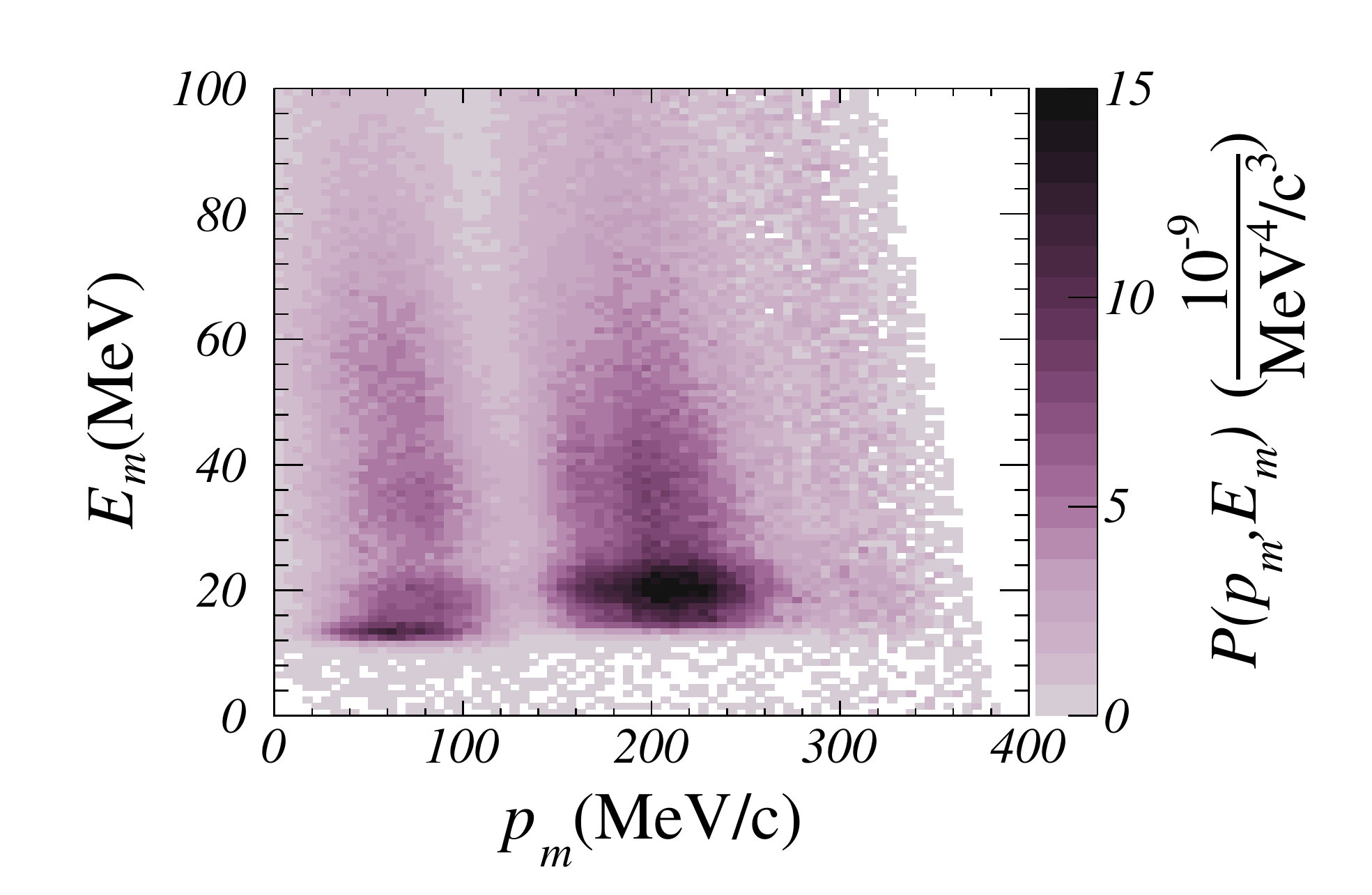}\label{fig:redxs_pm_Ar_a}} \\
\subfloat[MC]{\includegraphics[width=1.0\columnwidth]{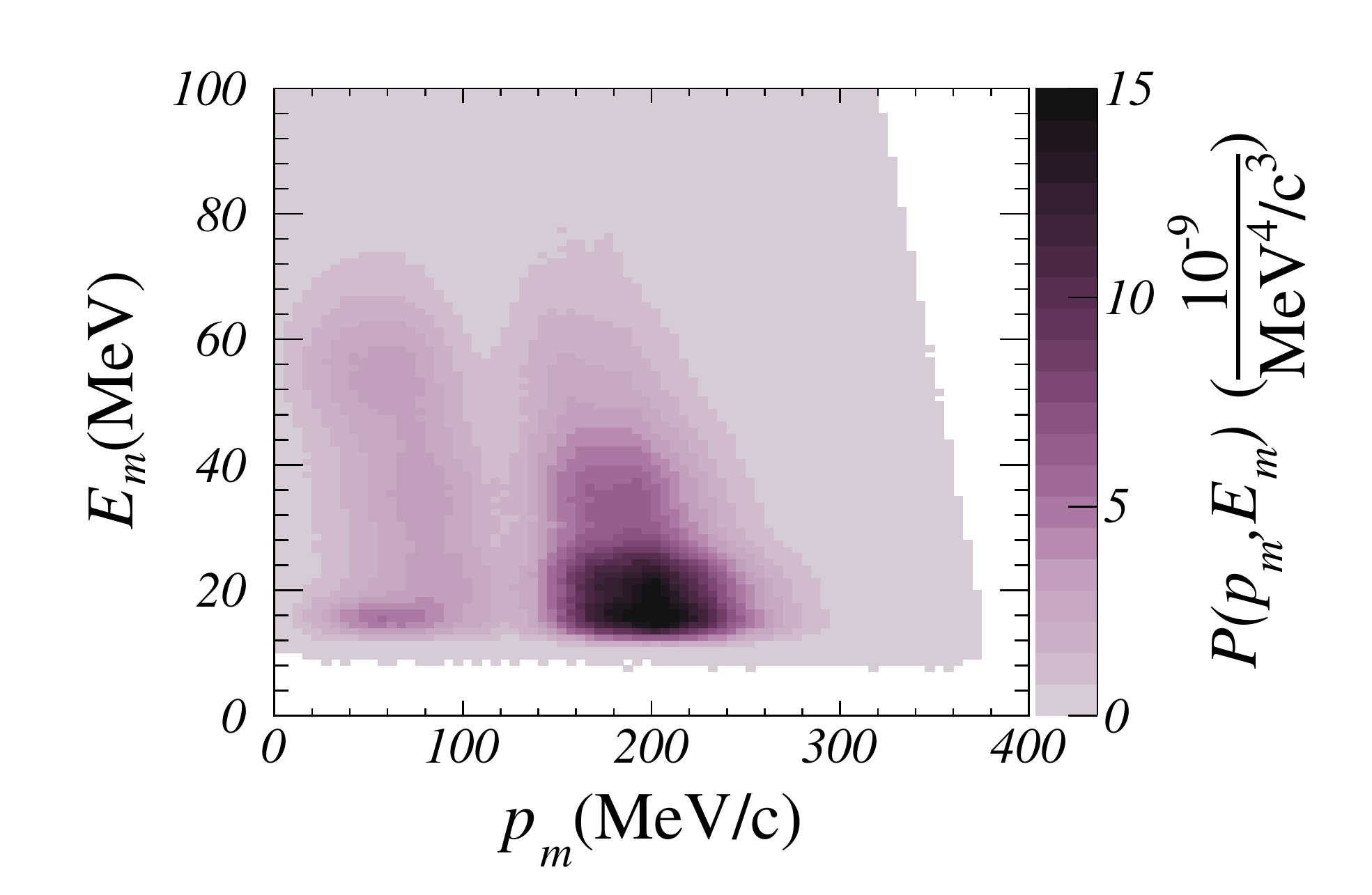}\label{fig:redxs_pm_Ar_b}}
\caption{Cross section as function of $p_m$ and $E_m$ obtained from (a) data and (b) Monte Carlo simulation for all the kinematics. The simulation is corrected for final-state interactions. }\label{fig:redxs_pm_Ar}
\end{figure}

\section{Reduced cross section comparison}\label{sec:comparison}
%
%%%%%%%%%%%%%%%%%%%%%%%%%%%%%%%%%%%%%%%%%%%%%%%%%%%%%%%%%%

Figure~\ref{fig:redxs_pm_Ar} shows the reduced cross section for both data and MC\textemdash obtained using the cc1 prescription of Ref.~\cite{DeForest:1983ahx} for the off-shell $ep$ cross section\textemdash as a function of missing energy and missing momentum. Data events have been background subtracted following the methods described in our previous paper~\cite{Gu:2020rcp}, while MC events have been corrected for FSI. 

As shown in Fig.~\ref{fig:redxs_pm_Ar_b}, the MC simulation exhibits a discontinuity at $E_m\sim 20$~MeV. Its origin can be traced back to the simple treatment of the FSI correction~\cite{Gu:2020rcp}, which shifts the cross section in the $p_m$ direction by a constant depending on the dominant shell. When the dominant shell changes, a discontinuity occurs. The dominant shells are identified according to the preliminary estimates of the peak positions and widths. The differences between the preliminary values and those extracted from the analysis are treated as systematic uncertainties in the DWIA correction of our MC. We have studied these effects by varying the DWIA correction in the MC and repeating the spectroscopic analysis described in Sec.~\ref{sec:fit}, and we did not see any appreciable variations of the fit results.

%%%%%%%%%%%%%%%%%%%%%%%%%%%%%%%%%%%%%%%%%%%%%%%%%%%%%%%%%%%%%%
\section{Spectroscopic analysis}\label{sec:fit}

As it is apparent in Fig.~\ref{fig:missing_energy_distribution}, the missing-energy spectrum of \isotope[40][]{Ar} is expected to exhibit three main peaks. In our data analysis, the $E_m$ ranges of 0 to 30 MeV, 30 to 54 MeV, and 54 to 90 MeV are treated separately. To increase the sensitivity, the fit of the missing momentum distribution is broken down into three contributions associated with the three missing energy regions.

The expected missing momentum distribution integrated over the full spectrum of missing energy is displayed in Fig.~\ref{fig:missing_momentum_contributions_for_setups}, together with reference points identifying our five different kinematic settings. 

We have performed a fit to the experimental missing energy and missing momentum distributions to extract spectroscopic factors, mean value and width of each of the \isotope[40][]{Ar} orbitals. 

For each bin in the spectra of missing energy (100 bins between 1 and 100~MeV) and missing momentum (40 bins with momentum range changing between kinematics), we determined the product of the reduced MC cross section~\cite{Arrington:1998ps} and the ratio of the data to simulation yield, 
\begin{equation}
 \frac{d^{2}\sigma^\text{red}_{cc1}}{d\Omega dE'} = \Big(\frac{d^{2}\sigma^\text{red}_{cc1}}{d\Omega dE'}\Big)_{\text{MC}} \times \frac{Y(E',\theta)}{Y_{\text{MC}}(E',\theta)},
 \label{eq:yield_for_chi2}
\end{equation}
where the $Y(E',\theta)$ is the yield for a given bin and the reduced MC cross section is a fit to the existing data~\cite{SIMC}. The reduced cross section includes (i) the $\sigma_{cc1}$ cross section of de~Forest~\cite{DeForest:1983ahx}, (ii) the predictions of the SF model, (iii) radiative corrections~\cite{Mo:1968cg},  (iv) Coulomb corrections~\cite{Aste:2005wc}, (v) changes in radiation length of the target due to the target-boiling effect~\cite{Santiesteban:2018qwi,Dai:2018xhi} and  (vi) DWIA corrections.

The fit performs a $\chi^2$ minimization using the {\sc minuit}~\cite{James:2004vla} package available in {\sc root}~\cite{ROOT}. The missing energy
distributions of the shell-model states are assumed to follow the Maxwell-Boltzmann distribution
\begin{equation}\label{eq:maxwell}
\begin{split}
F_\alpha(E_m) &= \frac{4}{\sqrt{\pi}\sigma} \left ( \frac{E_m-E_\alpha+\sigma_\alpha}{\sigma_\alpha} \right )^2\\
&\qquad\times\exp \left [-\left (\frac{E_m-E_\alpha+\sigma_\alpha}{\sigma_\alpha}\right )^2 \right ],
\end{split}
\end{equation}
where $E_\alpha$ denotes the position of the peak, the width of which is determined by $\sigma_\alpha$. The full width at half maximum (FWHM), is given by $\Gamma_\alpha=1.1549\sigma_\alpha$. We require that $E_m\geq E_\alpha-\sigma_\alpha$.

\par We have employed  the $\chi^2$ function:
\begin{equation}
\chi^2 = \sum_i\chi^2_i = \sum_i\left ( \frac{ \sigma_i^\text{red, obs} - \sum_\alpha{S_\alpha f_\alpha^\text{pred}(i)} }{\sigma_{\sigma_i^\text{red, obs}}} \right )^2,
\label{eq:chi2_i}
\end{equation}
where the index $i$ labels the missing momentum bin, $\alpha$ is the orbital index, $f_\alpha^\text{pred}(i)$ is the parametrized prediction evaluated at bin $i$ in the missing momentum spectra for orbital $\alpha$, $S_\alpha$ is the spectroscopic factor. 

The missing momentum distribution does not show dependence on the mean energies and widths of the orbitals. Uncertainties in the fit parameters have been increased during the fit minimization to make the final $\chi^2$/d.o.f. equal to 1.

We used data from our five kinematic settings and for each kinematics we integrated three different missing energy regions, as discussed above. 

The performed fits cover the $p_m$ range specified in Table~\ref{tab:ar_S_contributions} and the $E_m$ values between 10 and 70 MeV, covering $\sim$90\% and $\sim$41\% of the mean-field and correlated parts of the spectral function, respectively.

In the case of the minimization in missing momentum, the results are summarized in Table~\ref{tab:momentum_fit_results}, where we report all the results including degrees of freedom and the value of the $\chi^2$. We also repeated the fit excluding the correlated SF contribution.

The spectroscopic factors reported in Tables~\ref{tab:momentum_fit_results} and \ref{tab:momentum_energy_spectra} are normalized to $0.8 \times 18$ for the total strength of the orbitals and to $0.2 \times 18$ for the correlated part, including the corrections on phase space coverage.

\begin{table}
\centering
\caption{\label{tab:momentum_fit_results} Comparison of the results of the $\chi^2$ minimization using the missing momentum distributions, obtained with and without the correlated spectral function. For every state $\alpha$, we present the extracted spectroscopic factor $S_\alpha$, and its occupation number in the independent-particle shell model, $N_\alpha$. Additionally, we provide the total spectroscopic strength, the number of degrees of freedom (d.o.f.), and the $\chi^2$ per d.o.f.}
\begin{ruledtabular}
\begin{tabular}{c c r r }
   & &  \multicolumn{1}{c}{w/ corr.}      & \multicolumn{1}{c}{w/o corr.} \\
\colrule
$\alpha$    & $N_\alpha$   & \multicolumn{2}{c}{$S_\alpha$}  	\\
\colrule
$1d_{3/2}$  & 2      & $0.78\pm0.05$ & \phantom{0}$0.78\pm0.09$   \\
$2s_{1/2}$  & 2         & $2.07\pm0.07$	& \phantom{0}$2.10\pm0.10$  \\
$1d_{5/2}$  & 6	        & $2.27\pm0.04$	& \phantom{0}$2.27\pm0.08$  \\
$1p_{1/2}$  & 2	        & $2.72\pm1.23$	& \phantom{0}$2.72\pm0.34$  \\
$1p_{3/2}$  & 4	        & $3.36\pm0.04$	& \phantom{0}$3.53\pm0.06$  \\
$1s_{1/2}$  & 2	        & $2.54\pm0.04$	& \phantom{0}$2.65\pm0.02$  \\
corr.       & 0       & $0.48\pm0.01$  & \multicolumn{1}{c}{excluded}\\
\colrule
$\sum_\alpha S_\alpha$  &      & $14.48\pm1.24$          & $14.05\pm0.38$ \\
d.o.f. &   & 1,132         & 1,133  \\
$\chi^2$/d.o.f.&      & 1.9           & 3.2                       \\
\end{tabular}
\end{ruledtabular}
\end{table}

We then minimized the $\chi^2$ function using the missing energy spectra,
\begin{equation}
\chi^2 = \sum_i{\chi^2_i} + \sum_n{\left (\frac{\tau_n^\textrm{fit} - \tau_n^c}{\sigma_n^\textrm{fit}}\right)^2},
\label{eq:chi2_constraints}
\end{equation}
including constraints on the position of the topmost energy levels from previous experimental results~\cite{Chen:2018trb,Wang:2017,Mairle:1993asu,Mairle:1993ntp}, summarized in Table~\ref{tab:Constraints}.

\begin{table}[tb]
\centering
\caption{\label{tab:Constraints} Constraints on the fits to the missing-energy spectra obtained from past measurements~\cite{Chen:2018trb,Wang:2017,Mairle:1993asu,Mairle:1993ntp}. For the clarity of presentation, we denote $E_\alpha$ as $E(\alpha)$.}
\begin{ruledtabular}
\begin{tabular}{c  d  d }
Parameter     		    & \multicolumn{1}{c}{Value (MeV)}  	    & \multicolumn{1}{c}{Uncertainty (MeV)} 	\\
\colrule
$E(1d_{3/2})$ & 12.529 & 0.002 \\
$E(2s_{1/2})$ & 12.925 & 0.002 \\
$E(1d_{5/2})$ & 18.229 & 0.015 \\
$E(1p_{3/2})-E(1p_{1/2})$  & 4.1  & 1.5 \\
\end{tabular}
\end{ruledtabular}
\end{table}

\begin{table}
\centering
\caption{\label{tab:momentum_energy_spectra} Comparison of the results of the $\chi^2$ minimization using the missing energy distributions, obtained with all priors, without priors from the missing-momentum fits, and without the correlated spectral function. For every state $\alpha$, we present the extracted spectroscopic factor $S_\alpha$, and its occupation number in the independent-particle shell model, $N_\alpha$. Additionally, we provide the total spectroscopic strength, the number of degrees of freedom (d.o.f.), and the $\chi^2$ per d.o.f.}
\begin{ruledtabular}
\begin{tabular}{c c r r r } 
& & \multicolumn{1}{c}{all priors}     & \multicolumn{1}{c}{w/o $p_m$}      & \multicolumn{1}{c}{w/o corr.} \\
\colrule
$\alpha$    & $N_\alpha$    & \multicolumn{3}{c}{$S_\alpha$} \\      
\colrule
$1d_{3/2}$  & 2         & $0.89\pm0.11$ & $1.42\pm0.20$  & \phantom{0}$0.95\pm0.11$ \\
$2s_{1/2}$  & 2         & $1.72\pm0.15$	& $1.22\pm0.12$  & \phantom{0}$1.80\pm0.16$ \\
$1d_{5/2}$  & 6	        & $3.52\pm0.26$	& $3.83\pm0.30$  & \phantom{0}$3.89\pm0.30$ \\ 
$1p_{1/2}$  & 2	        & $1.53\pm0.21$	& $2.01\pm0.22$	 & \phantom{0}$1.83\pm0.21$ \\
$1p_{3/2}$  & 4	        & $3.07\pm0.05$	& $2.23\pm0.12$	 & \phantom{0}$3.12\pm0.05$ \\
$1s_{1/2}$  & 2	        & $2.51\pm0.05$	& $2.05\pm0.23$  & \phantom{0}$2.52\pm0.05$\\
corr.  	    & 0         & $3.77\pm0.28$ & $3.85\pm0.25$  & \multicolumn{1}{c}{excluded}	\\ 
\colrule
$\sum_\alpha S_\alpha$ & & $17.02\pm0.48$         & $16.61\pm0.57$           & $14.12\pm0.42$ \\
d.o.f       &           & 206           & 231            & 232 \\
$\chi^2$/d.o.f.&      & 1.9           & 1.4            & 2.0 \\
\end{tabular}
\end{ruledtabular}
\end{table}

The spin-orbit splitting has been computed using the phenomenological prescription of Ref.~\cite{Mairle:1993asu,Mairle:1993ntp},
\begin{equation}
    E(n,l,l-1/2)-E(n,l,l+1/2) = \frac{2 l + 1}{2n} k A^{-C},
    \label{eq:SO_theory}
\end{equation}
with angular momentum $l$, main quantum number $n$, and mass number $A$. The empirically determined constants $k=23.27$~MeV and $C=0.583$~\cite{Mairle:1993asu} are included in the fit as penalty function to the $\chi^2$. The uncertainty value has been calculated comparing the predictions of Eq.~\eqref{eq:SO_theory} with the available experimental data from NIKHEF-K~\cite{Kramer:1989uiu,Kramer:1990,Leuschner:1994zz}. We apply this constraint only to the $1p$ shells.

The missing energy spectra minimization returns 20 parameters: 3 parameters for each orbital (the spectroscopic factor, the position of the maximum, and the width of the distribution) and 2 parameters for the correlated SF (the strength and the threshold energy). 
We present our results in Table~\ref{tab:momentum_energy_spectra}. We repeated the fit excluding the results coming from the $p_m$ minimization and without the correlated SF part. 

All the results are compatible within errors, which indicates no large bias in the determination of the spectroscopic factors using different set of priors. 

\begin{table}
\centering
\caption{\label{tab:energy_spectra_frpm_fits} The peak positions $E_\alpha$, their widths $\sigma_\alpha$, and the parameter $E_\text{corr}$ of the correlated spectral function obtained from the $\chi^2$ minimization of missing energy distributions. The results with and without priors from the missing momentum fit are compared.}
\begin{ruledtabular}
\begin{tabular}{c r r r r }
    & \multicolumn{2}{c}{$E_\alpha$ (MeV)}  	& \multicolumn{2}{c}{$\sigma_\alpha$ (MeV)} \\ 
\cline{2-3}\cline{4-5}\\[-8pt]
$\alpha$    & \multicolumn{1}{c}{w/ priors}     & \multicolumn{1}{c}{w/o priors}    & \multicolumn{1}{c}{w/ priors}        & \multicolumn{1}{c}{w/o priors} \\      
\colrule
$1d_{3/2}$  & $12.53\pm0.02$                        & $10.90\pm0.12$    & \phantom{0}$1.9\pm0.4$    & \phantom{0}$1.6\pm0.4$ \\
$2s_{1/2}$ 	& $12.92\pm0.02$                        & $12.57\pm0.38$    & \phantom{0}$3.8\pm0.8$    & \phantom{0}$3.0\pm1.8$ \\
$1d_{5/2}$	& $18.23\pm0.02$                        & $17.77\pm0.80$    & \phantom{0}$9.2\pm0.9$    & $9.6\pm1.3$\\ 
$1p_{1/2}$	& $28.8\phantom{0}\pm0.7\phantom{0}$	& $28.7\phantom{0}\pm0.7\phantom{0}$	& $12.1\pm1.0$              & \phantom{0}$12.0\pm3.6$\\
$1p_{3/2}$	& $33.0\phantom{0}\pm0.3\phantom{0}$	& $33.0\phantom{0}\pm0.3\phantom{0}$	& \phantom{0}$9.3\pm0.5$    & \phantom{0}$9.3\pm0.5$ \\
$1s_{1/2}$	& $53.4\phantom{0}\pm1.1\phantom{0}$	& $53.4\phantom{0}\pm1.0\phantom{0}$ 	& $28.3\pm2.2$              & \phantom{0}$28.1\pm2.3$ \\
corr.  		& $24.1\phantom{0}\pm2.7\phantom{0}$    & $24.1\phantom{0}\pm1.7\phantom{0}$	    & \multicolumn{1}{c}{---}	& \multicolumn{1}{c}{---} \\
\end{tabular}
\end{ruledtabular}
\end{table}

%%%%%%%%%%%%%%%%%%%%%%%%%%%%%%%%%%%%%%%%%%%%%%%%%%
\begin{figure}
\centering
\subfloat[$\quad15 <p_m < 110$~MeV/c]{\includegraphics[width=0.9\columnwidth]{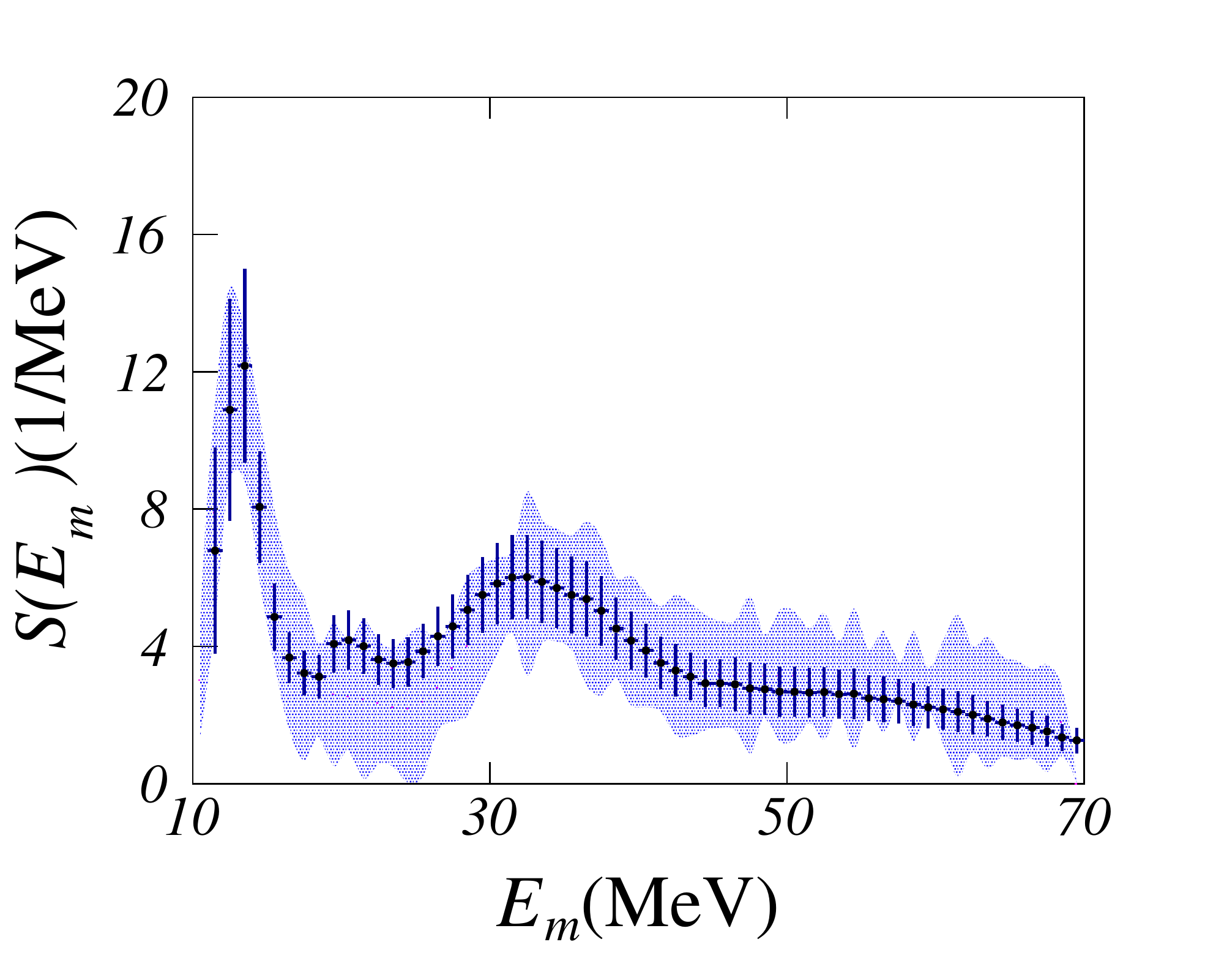}}\\
\subfloat[$\quad140 <p_m < 210$~MeV/c]{\includegraphics[width=0.9\columnwidth]{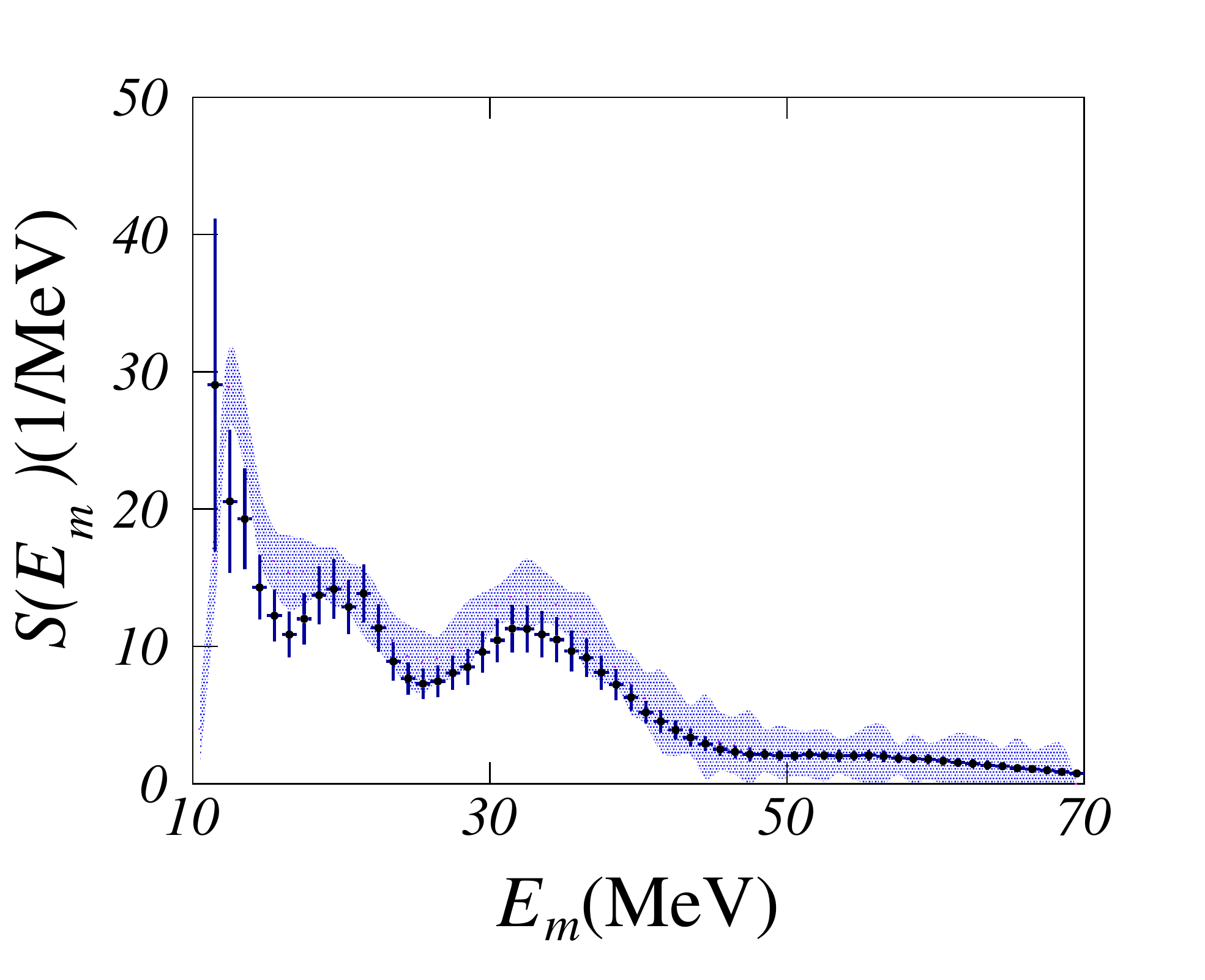}}\\
\subfloat[$\quad120 <p_m < 220$~MeV/c]{\includegraphics[width=0.9\columnwidth]{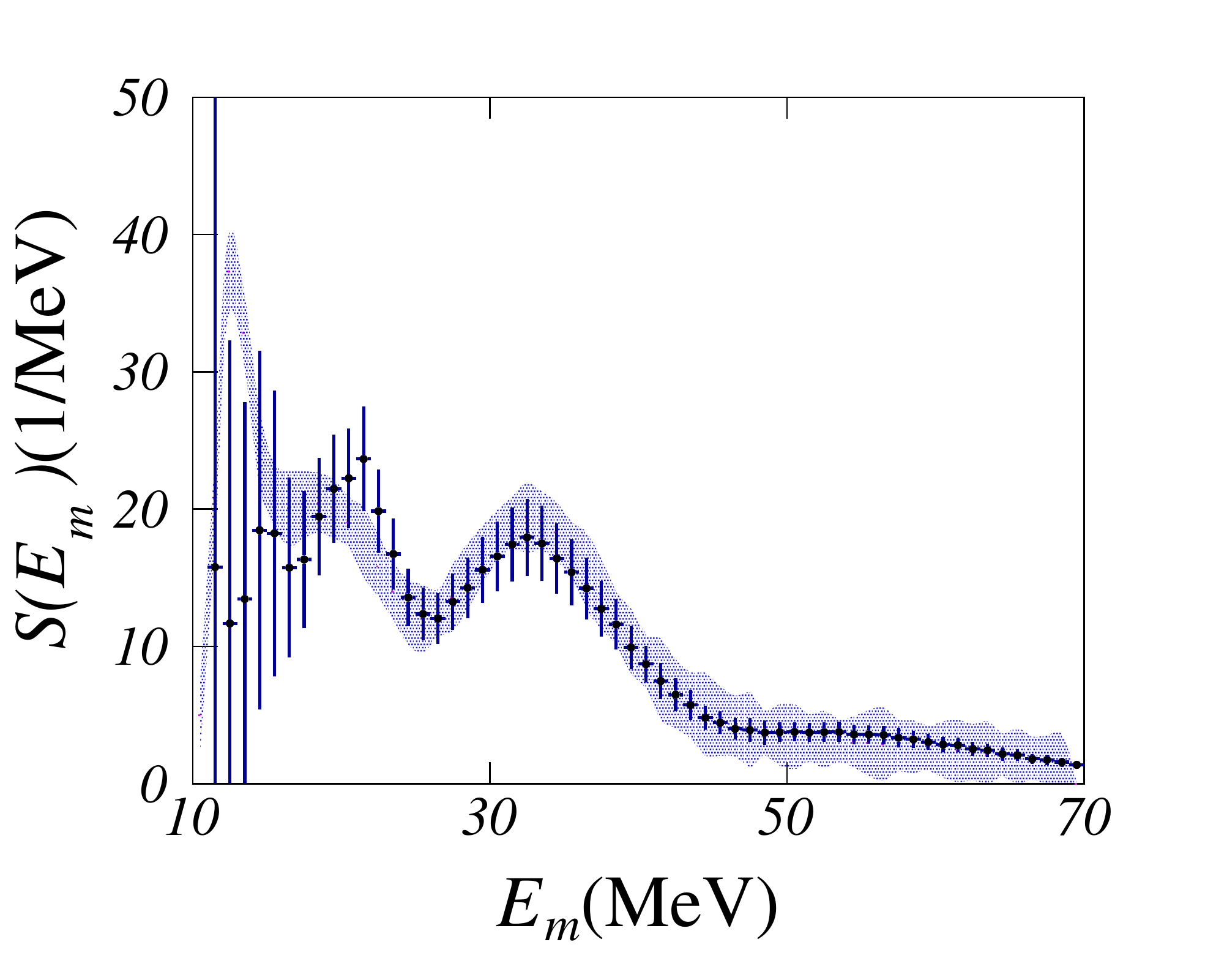}}\\
\caption{\label{fig:missing_energy_spectra1} Missing energy distributions obtained for the kinematic settings of Table~\ref{tab:kinematic}. The blue band shows the results of our fit including the full error budget. }
\end{figure}

%%%%%%%%%%%%%%%%%%%%%%%%%%%%%%%%%%%%%%%%%%%%%%%%%%
\begin{figure}
\centering
\subfloat[$\quad190 <p_m < 250$~MeV/c]{\includegraphics[width=0.9\columnwidth]{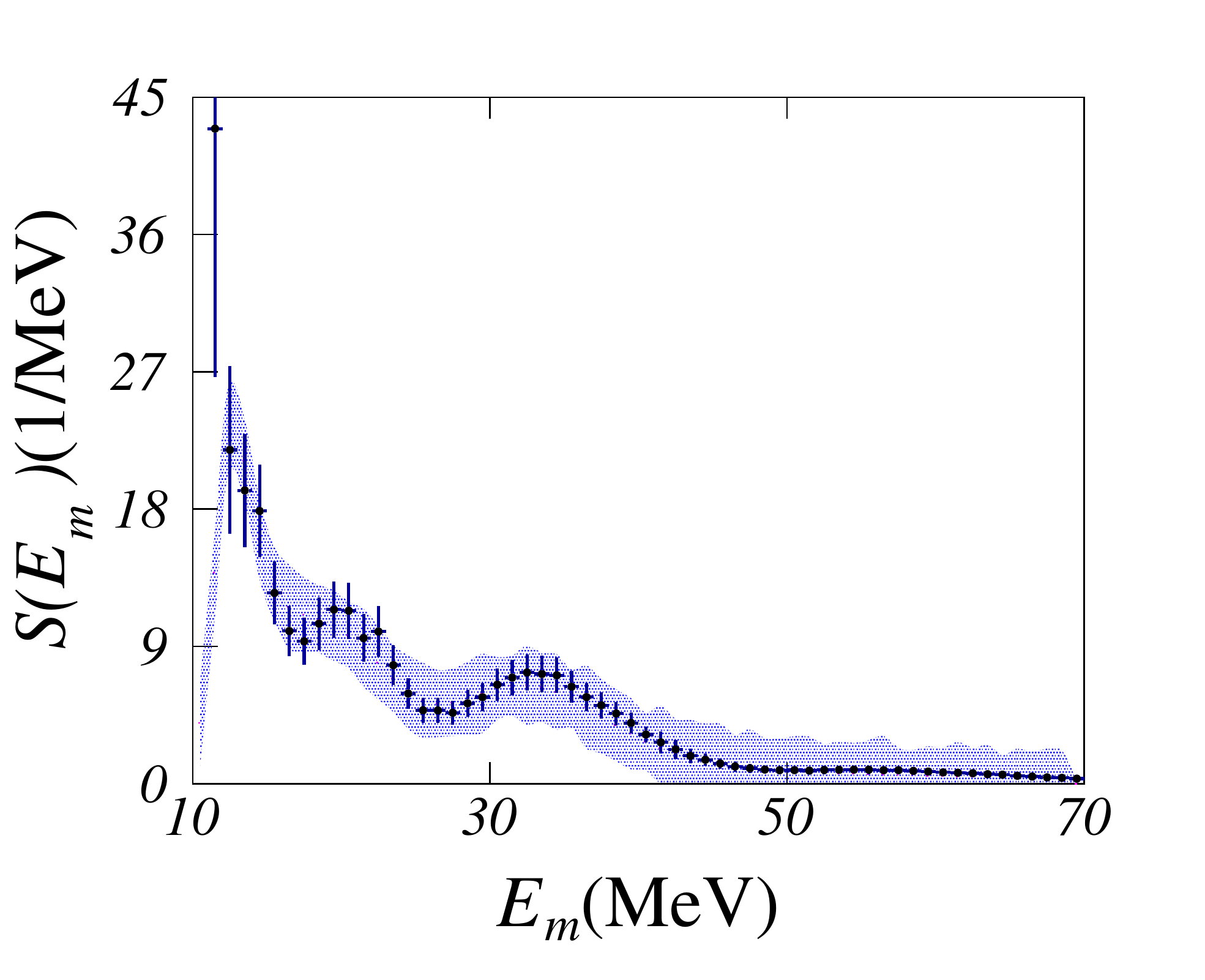}}\qquad
\subfloat[$\quad260 <p_m < 320$~MeV/c]{\includegraphics[width=0.9\columnwidth]{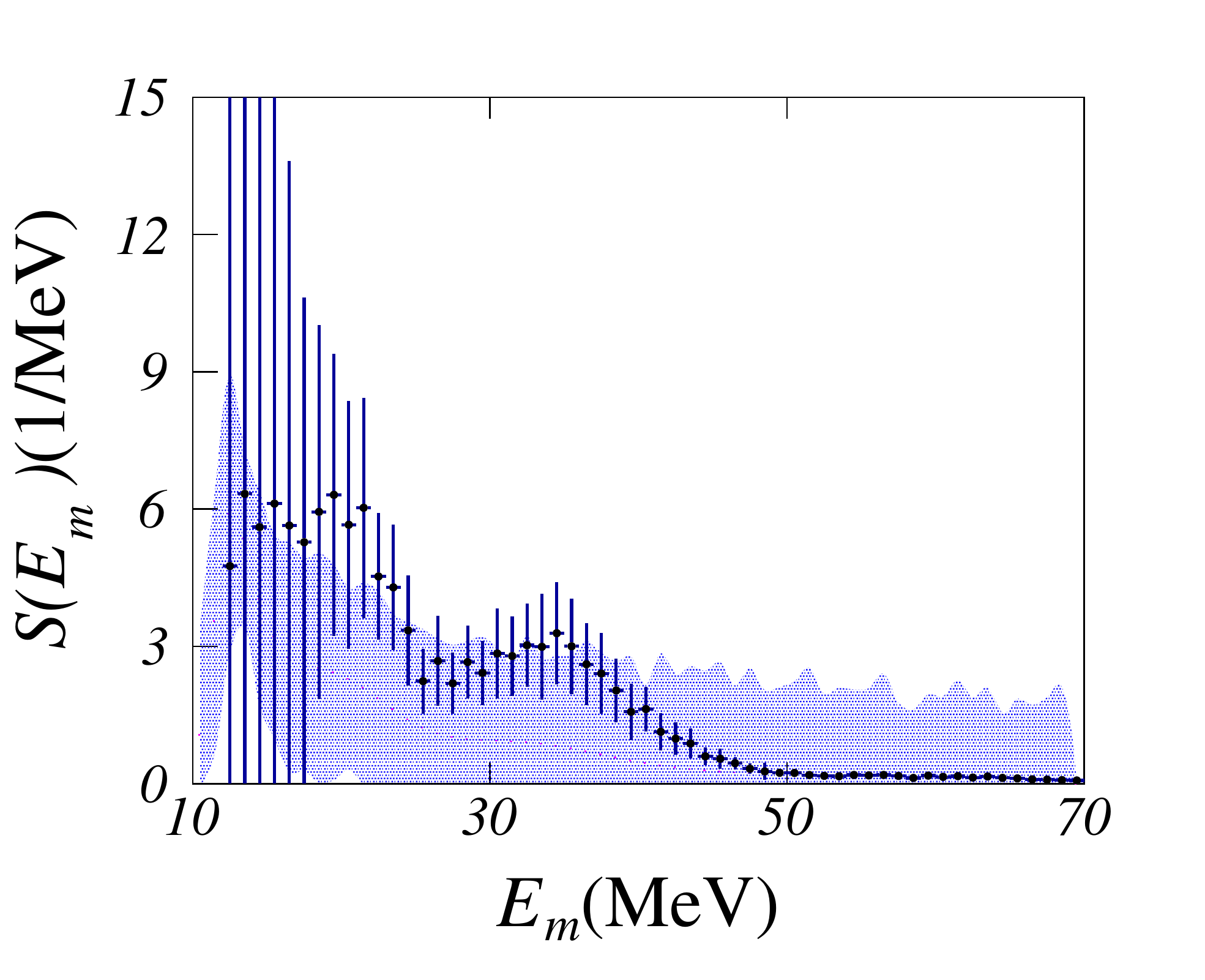}}\qquad
\caption{\label{fig:missing_energy_spectra2} Continued from Fig.~\ref{fig:missing_energy_spectra1}. }
\end{figure}

\begin{figure}[bt]
    \centering
\subfloat[$\quad0 <E_m < 30$~MeV]{\includegraphics[width=0.9\columnwidth]{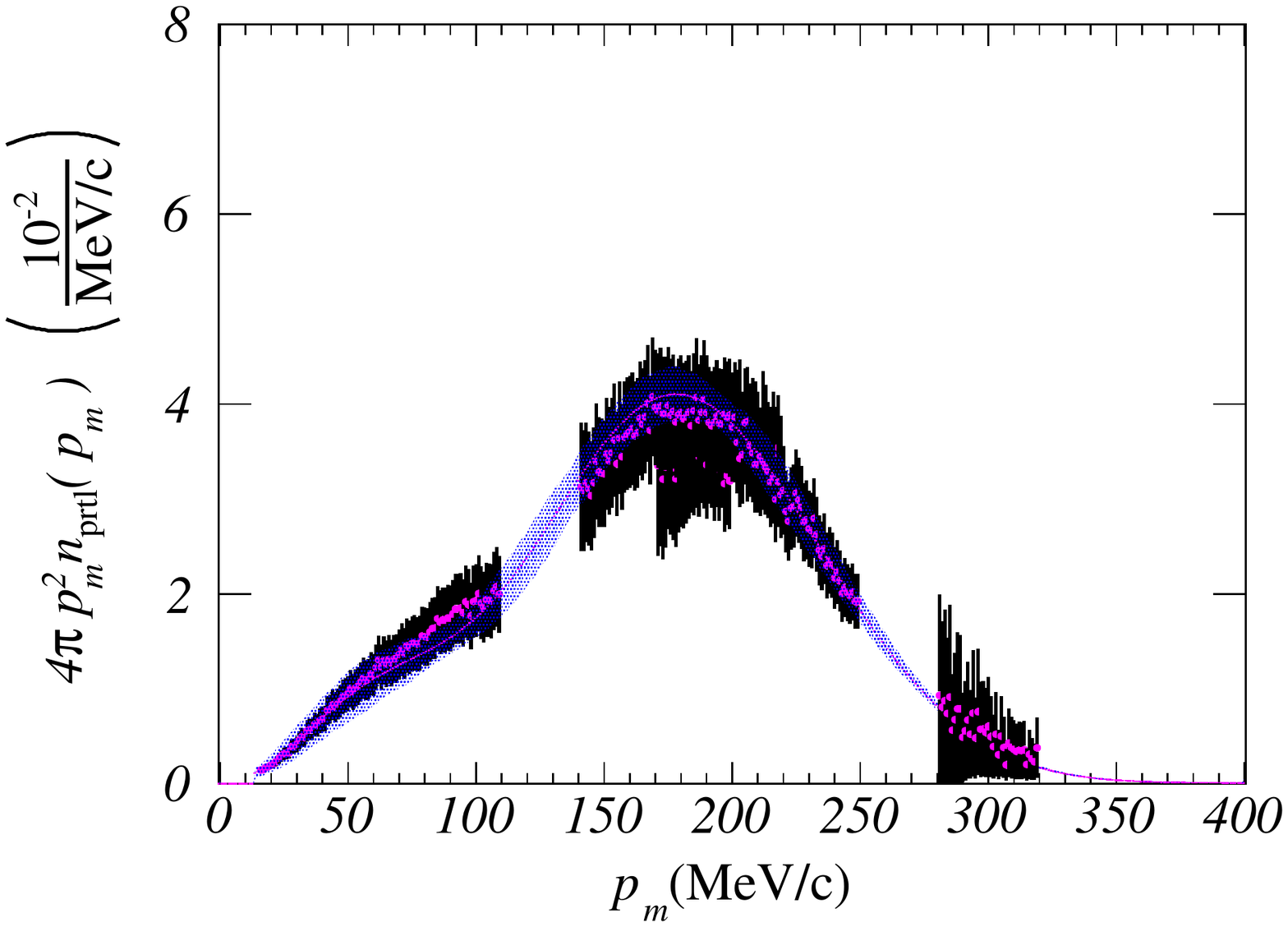}}\qquad
\subfloat[$\quad30 < E_m < 54$~MeV]{\includegraphics[width=0.9\columnwidth]{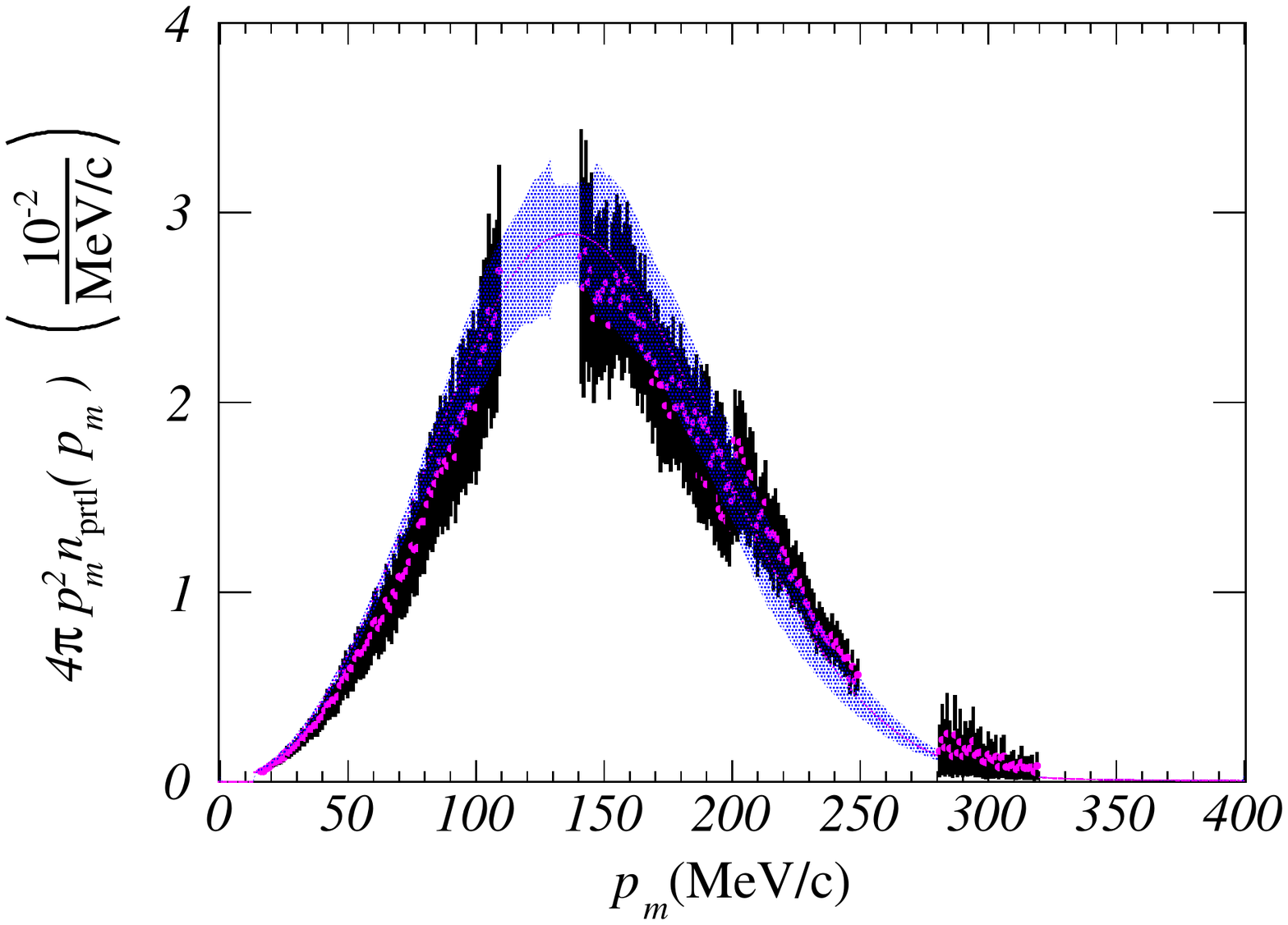}}\qquad
\subfloat[$\quad54 <E_m < 90$~MeV]{\includegraphics[width=0.9\columnwidth]{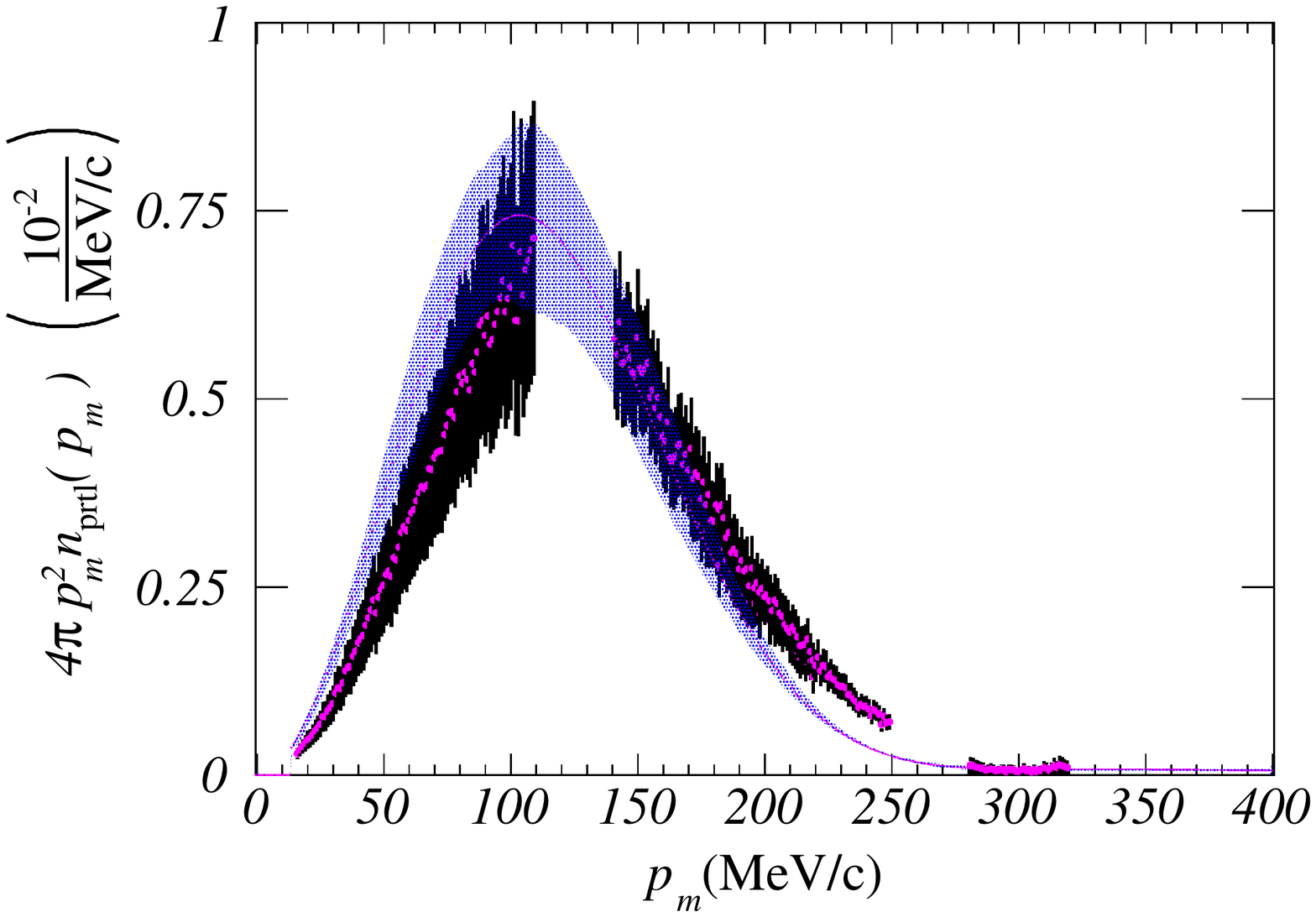}}\qquad
       \caption{Partial momentum distribution obtained by integrating the test spectral function over the missing energy range of (a) 0--30 MeV, (b) 30--54 MeV, and (c) 54--90 MeV, presented with the geometric factor of $4\pi p_m^2$.}
\label{fig:missing_momentum_fit_results}
\end{figure}

\begin{figure*}[bt]
    \centering
\includegraphics[width=0.8\textwidth]{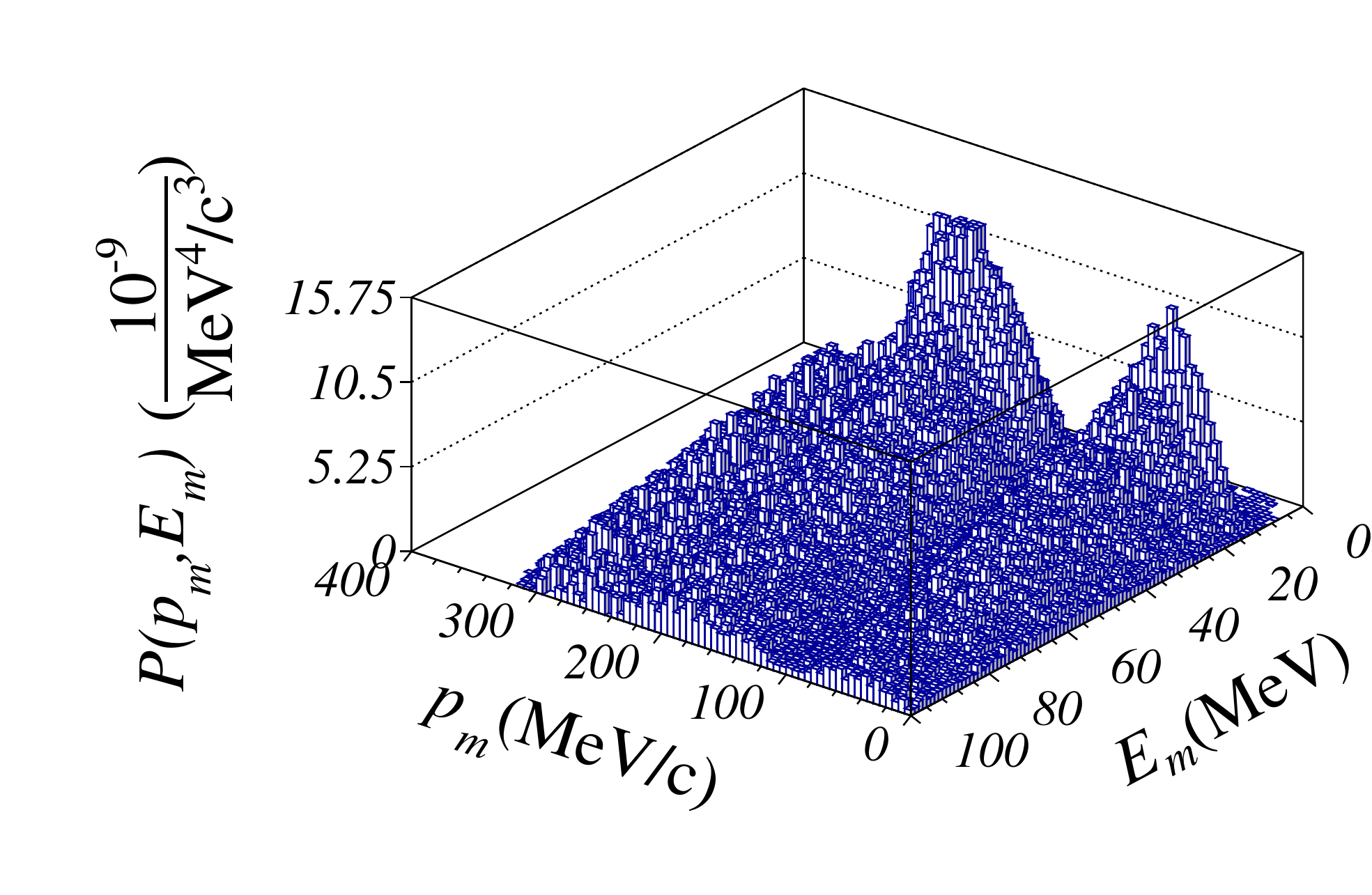}
\caption{Reduced cross section as function of missing energy and missing momentum.}
\label{fig:3D_momentum_energy}
\end{figure*}

We do not see a large bias introduced by the set of priors or the theory model that we use, but clearly the fit with the correlated SF is a better representation of our data.

We have also repeated the minimizations using different sets of priors for the orbital parametrizations: the Maxwell-Boltzmann or Gaussian distributions, with the width governed by a constant or linearly dependent on the distance from the Fermi energy, $E_m-E_F$. The results obtained are all compatible within errors, which indicates that the fit is relatively independent of the parametrisation used.

The results of Figs.~\ref{fig:missing_energy_spectra1} and~\ref{fig:missing_energy_spectra2} show that the test spectral function model, rescaled using the parameters obtained from the 
fit, listed in Table~\ref{tab:momentum_energy_spectra}, is capable to reproduce our data quite accurately. Figure~\ref{fig:missing_momentum_fit_results} reports the missing momentum distributions obtained integrating the data and the model over three different ranges of missing energies: 0--30~MeV, 30--54~MeV and 54--90~MeV. In Figs.~\ref{fig:missing_energy_spectra1}--\ref{fig:missing_momentum_fit_results}, the SF model obtained from the fit  is represented by the blue band, which accounts for the full error budget, including correlations and FSI correction uncertainties. The experimental coverage in the missing momentum is not complete due to experimental conditions and beam time limitation, and the fact that in kinematics 4 and 5 we were not able to collect all the desired data. In kinematics 4 and 5, the data that we have in this $p_m$ region is statistically limited.

In the case of the missing momentum distributions, it is apparent that our experimental data cover the relevant kinematic range with just a few exceptions,  most notably at vanishing $p_m$. A remarkable feature of Fig.~\ref{fig:missing_momentum_fit_results} is the agreement, within admittedly sizable uncertainties, of the reduced cross sections corresponding to kinematics 2 and 3, see Table~\ref{tab:kinematic}. These kinematics, while spanning similar $p_m$ and $E_m$ ranges, correspond to different electron scattering angles and energy transfers, and to different energies and emission angles of the outgoing proton. Therefore, the observed 
agreement appears to suggest the reliability of DWIA treatment of FSI\textemdash the effects of which are different in the two kinematic settings\textemdash
and, more generally, the validity of the factorisation scheme underlying our analysis.

For the sake of completeness, in Fig.~\ref{fig:3D_momentum_energy}, we also provide a three-dimensional representation of the data, displayed as a function of missing energy and missing momentum.

%%%%%%%%%%%%%%%%%%%%%%%%%%%%%%%%%%%%%%%%%%%%%%%%%%%%%%%%%%%%%%
\section{Comparison to previous measurements for \texorpdfstring{\ensuremath{\isotope[40]{Ar}}}{Ar} and \texorpdfstring{\ensuremath{\isotope[40]{Ca}}}{Ca}} \label{sec:comparisonToPast}
%%%%%%%%%%%%%%%%%%%%%%%%%%%%%%%%%%%%%%%%%%%%%%%%%%%%%%%%%%%%%%

\begin{table}[tb]
\centering
\caption{\label{table:Ar&Ca_energy} Comparison of the peak positions $E_\alpha$ of the shell-model states of protons in the argon and calcium nuclei. All values are given in MeV. }
\begin{ruledtabular}
\begin{tabular}{lcc}
\multirow{2}{*}{$\alpha$} & \multicolumn{1}{c}{\isotope{Ar}$(e,e'p)$} & \multicolumn{1}{c}{\isotope{Ca}$(p,2p)$}  \\
 & \multicolumn{1}{c}{this analysis} & \multicolumn{1}{c}{Ref.~\cite{Volkov:1990rs}}\\
\colrule
$1d_{3/2}$ & $12.53\pm0.02$  & $\phantom{0}8.5\pm0.1$\\
$2s_{1/2}$ & $12.92\pm0.02$  & $11.0\pm0.1$\\
$1d_{5/2}$ & $18.23\pm0.02$  & $15.7\pm0.1$\\
$1p_{1/2}$ & $28.8\phantom{0}\pm0.7\phantom{0}$  &  $29.8\pm0.7$\\
$1p_{3/2}$ & $33.0\phantom{0}\pm0.3\phantom{0}$  & $34.7\pm0.3$\\
$1s_{1/2}$ & $53.3\phantom{0}\pm1.1$\phantom{0}  & $53.6\pm0.6$ \\
\end{tabular}
\end{ruledtabular}
\end{table}

In Table~\ref{table:Ar&Ca_energy}, we compare the proton spectrum of \isotope[40][18]{Ar} resulting from our analysis with that of \isotope[40][20]{Ca}, as reported by Volkov \etal~\cite{Volkov:1990rs}. The calcium results were obtained performing a coincidence $(p,2p)$ experiment with the 1-GeV proton beam delivered by the Gatchina synchrocyclotron. 
The observed missing-energy spectra were decomposed as a sum of Gaussian distributions and a uniform background.

It has to be pointed out that, unlike \isotope[40][18]{Ar}, \isotope[40][20]{Ca} is a symmetric and closed-shell nucleus. These features, and the different number of protons, introduce significant differences between the valence shells\footnote{Our estimates of the $1d_{3/2}$ and $2s_{1/2}$ peak positions for \isotope[36][18]{Ar} (\isotope[40][20]{Ca}) are 8.51 and 9.73 MeV~\cite{Chen:2018trb,Wang:2017,Mairle:1993asu} (8.33 and 10.85 MeV~\cite{Wang:2017,Chen:2011dsm,Doll:1976qwt}), respectively}. However, as shown in Table~\ref{table:Ar&Ca_energy}, for the deeply bound states the peak positions turn out to agree to $\sim$2 MeV or better. Therefore, it is interesting to investigate the extent to which these two nuclei are similar in the context of spectroscopic factors. 

As stated in the introduction, electron-induced proton knockout allows to probe the whole nuclear volume, which is essential for an accurate determination of spectroscopic factors.
On the other hand, knockout by hadronic probes\textemdash scattering mostly on the nuclear surface\textemdash has the advantage of higher cross sections. 

In the previous sections, we have discussed the first measurement of the proton spectrum in argon carried out using coincidence electron scattering. Here we compare these findings with previous measurements performed using both argon and calcium targets and different beams. We discuss the spectroscopic factors extracted by Kramer \etal~\cite{Kramer:1989uiu,Kramer:1990} and Yasuda \etal~\cite{Yasuda:2010zz,Yasuda:2012} for \isotope[40]{Ca}, as well as those by Mairle \etal~\cite{Mairle:1993asu} and Doll \etal~\cite{Wagner:1969rfq,Doll:1974kmd} for \isotope[40]{Ar}.

\begin{table*}[tb]
\centering
\caption{\label{table:Ar&Ca} Comparison of the occupation probabilities, $n(\alpha)=S_\alpha/N_\alpha$, extracted by single-proton knockout experiments with different probes for the argon and calcium targets. The results for deuteron scattering~\cite{Mairle:1993asu,Wagner:1969rfq,Doll:1974kmd} are rescaled to correct for inaccuracies of the wave functions, nonlocality of the optical potential, and finite range of interaction according to Ref.~\cite{Kramer:2000kc}. The measurements~\cite{Mairle:1993asu} covered excitation energies up to 9 MeV, compared to 7 MeV in Refs.~\cite{Wagner:1969rfq,Doll:1974kmd}.}
\begin{ruledtabular}
\begin{tabular}{cccccc}
\multirow{2}{*}{$\alpha$} & \isotope{Ar}$(e,e'p)$ &
\multicolumn{1}{c}{\isotope{Ca}$(e, e'p)$} &
\multicolumn{1}{c}{\isotope{Ca}$(p^\text{pol},2p)$}  & \multicolumn{1}{c}{\isotope{Ar}($d^\text{pol}$,\isotope[3]{He})} &
\multicolumn{1}{c}{\isotope{Ar}($d$,\isotope[3]{He})}  \\
 & \multicolumn{1}{r}{this analysis} &
\multicolumn{1}{c}{Refs.~\cite{Kramer:1989uiu,Kramer:1990}} &
\multicolumn{1}{c}{Ref.~\cite{Yasuda:2010zz} revised~\cite{Yasuda:2012}}  & \multicolumn{1}{c}{Ref.~\cite{Mairle:1993asu} rescaled~\cite{Kramer:2000kc}} &
\multicolumn{1}{c}{Refs.~\cite{Wagner:1969rfq,Doll:1974kmd} rescaled~\cite{Kramer:2000kc}}  \\
\colrule
$1d_{3/2}+2s_{1/2}$ & $0.65\pm0.05$ & $0.64\pm0.05$ & $0.61\pm0.04$\footnotemark[1]{} & $0.62\pm0.13$ & $0.66\pm0.14$  \\
$1d_{3/2}$ & $0.45\pm0.06$ & $0.65\pm0.07$ & $0.65\pm0.05$\phantom{\footnotemark[1]{}} & $0.72\pm0.22$ & $0.77\pm0.23$ \\
$2s_{1/2}$ & $0.86\pm0.07$ & $0.64\pm0.06$ & $0.53\pm0.04$\footnotemark[1]{} & $0.51\pm0.15$ & $0.56\pm0.17$  \\
$1d_{5/2}$ & $0.59\pm0.04$ & $0.83\pm0.05$ & $0.85\pm0.09$\phantom{\footnotemark[1]{}} & $0.78\pm0.23$ & $0.54\pm0.16$ \\
$1p_{3/2}+1p_{1/2}$ & $0.77\pm0.04$ &  & $0.49\pm0.07$\phantom{\footnotemark[1]{}} &  &  \\
$1s_{1/2}$ & $1.25\pm0.03$ &  & $0.89\pm0.09$\phantom{\footnotemark[1]{}} &  &  \\
\end{tabular}
\end{ruledtabular}
\footnotetext[1]{Should the $1f_{7/2}$ contribution not be separated, $n(2s_{1/2})$ would be $0.83\pm0.05$, and $n(1d_{3/2})+n(2s_{1/2})$ would increase to $0.71\pm0.04$.}
\end{table*}

Kramer \etal~\cite{Kramer:1989uiu,Kramer:1990} performed a coincidence electron-scattering experiment on \isotope[40]{Ca} at NIKHEF-K and extracted the spectral function for missing momenta 0--280 MeV$/c$ and excitation energies $E_x$ below 22 MeV, employing beam energies $\sim$340--440 MeV.

Yasuda \etal~\cite{Yasuda:2010zz,Yasuda:2012} measured coincidence scattering of a polarized 392-MeV proton beam off \isotope[40]{Ca} at RCNP, covering missing momenta  0--200 MeV$/c$ and excitation energies up to $\sim$80 MeV. 
In the final version of the analysis~\cite{Yasuda:2012}, the $1d_{3/2}$ spectroscopic factor was fixed to the value extracted by Kramer \etal~\cite{Kramer:1989uiu}, and other spectroscopic factors were determined with respect to it.

Mairle \etal~\cite{Mairle:1993asu} used a polarized 52-MeV deuteron beam from the Karlsruhe cyclotron to probe the \isotope[40][18]{Ar} nucleus through a $(d,\isotope[3]{He})$ proton pickup reaction, varying the scattering angle. From the extracted differential cross sections and analyzing powers, spins, parities, and spectroscopic factors were assigned to individual levels observed for $E_x<6$ MeV and to the broad $1d_{5/2}$ level, measured up to the excitation energy of 9 MeV.

An earlier experiment at Karlsruhe by Doll \etal~\cite{Doll:1974kmd}, repeating and reanalyzing that of Wagner \etal~\cite{Wagner:1969rfq}, used an unpolarized 52-MeV deuteron beam to measure the differential cross sections for proton pickup from \isotope[40]{Ar}, which were used to obtain the spectroscopic factors for excitation energies up to 7~MeV.

In Ref.~\cite{Kramer:2000kc}, Kramer \etal analyzed the differences between electron- and deuteron-scattering measurements in great detail. They pointed out that the $(d,\isotope[3]{He})$ process, unlike $(e,e'p)$, does not probe the whole radial regions of the wave functions, but only their exponential tails, which are very sensitive to the exact shape of the assumed nuclear potential. As a consequence, the spectroscopic factors determined in past deuteron experiments suffered from inaccuracy of the employed wave functions, and from not accounting for nonlocality of the optical potential and for finite range of interaction. The global analysis of Kramer \etal, performed for targets ranging from \isotope[12]{C} to \isotope[208]{Pb}, showed that the spectroscopic factors of valence states determined by $(d,\isotope[3]{He})$ experiments were overestimated on average by 50\%.
Additionally, the authors of Ref.~\cite{Kramer:2000kc} assigned uncertainties---not reported in the original works---to the rescaled spectroscopic factors, estimating them to be $\sim$30\%. Here, we apply the findings of Kramer \etal~\cite{Kramer:2000kc} to the results of Refs.~\cite{Mairle:1993asu,Wagner:1969rfq,Doll:1974kmd}.

In Table~\ref{table:Ar&Ca}, we show the occupation probabilities---the spectroscopic factors normalized by the occupation numbers predicted by the independent-particle shell model, $n(\alpha)=S_\alpha/N_\alpha$---comparing our results with those of Refs.~\cite{Kramer:1989uiu,Kramer:1990,Yasuda:2010zz,Yasuda:2012,Wagner:1969rfq,Doll:1974kmd,Mairle:1993asu}. Except for the study by Yasuda {\it et al.}~\cite{Yasuda:2010zz,Yasuda:2012}, only the valence shells were probed in these experiments. The results are presented as occupation probabilities, to allow for a straightforward comparisons between targets with different proton numbers. 

As our analysis cannot clearly discriminate the contributions of the $1d_{3/2}$ and $2s_{1/2}$ states\textemdash the peaks of which are separated by 0.4~MeV\textemdash in addition to the individual occupation probabilities we include the results 
for their sum. This representation shows that the measurements performed using electron, proton, and deuteron beams at different kinematic regimes exhibit a remarkable agreement. 

The individual occupation probabilities for the $1d_{3/2}$ and $2s_{1/2}$ shells agree within uncertainties between all the past results. However, the results of our analysis differ from them. Interestingly, all the hadronic-beam experiments suggest that the $1d_{3/2}$ shell is more occupied than the $2s_{1/2}$ one, our analysis finds the opposite, and the past $(e,e'p)$ data show the same occupation probabilities. 

For the $1d_{5/2}$ shell, it is important to note that while the earlier deuteron-scattering experiments~\cite{Wagner:1969rfq,Doll:1974kmd} covered excitation energies up to 7~MeV, Mairle \etal~\cite{Mairle:1993asu}
found a significant strength of this heavily fragmented state extending up to (at least) 9~MeV. As a consequence, the occupation probability reported by Refs.~\cite{Wagner:1969rfq,Doll:1974kmd} can only be treated as a lower bound. 

The results of other past experiments for the $1d_{5/2}$ shell are in good agreement within uncertainties, especially having in mind that they are integrated over different ranges of missing energy. They point toward an occupation probability larger than that of the $1d_{3/2}$ and $2s_{1/2}$ shells, as expected when the distance from the Fermi energy increases~\cite{Benhar:1990zz}. On the other hand the value obtained from our analysis is significantly smaller, albeit consistent with all the deuteron-scattering results within their large uncertainties. Intriguingly, the $n(1d_{5/2})/n(1d_{3/2})$ ratio is $1.3\pm0.2$ for all but the deuteron-scattering experiments, and $1.1\pm0.6$ for the results of Mairle \etal~\cite{Mairle:1993asu}. The source of this deviation from the independent-particle shell-model expectation of 1 remains to be investigated in the future analyses with reduced uncertainties.

Because Yasuda \etal~\cite{Yasuda:2010zz,Yasuda:2012} did not resolve the 1/2 and 3/2 states of the $1p$ shell, we also combine them in Table~\ref{table:Ar&Ca}. As expected for a deeply bound state~\cite{Benhar:1990zz}, our result does not show a large depletion. It is, therefore, quite surprising to find that the proton-scattering experiment reports a significantly lower occupation probability. 

The spectroscopic factor of the $1s_{1/2}$ orbital obtained from the analysis presented here exceeds the prediction of the independent-particle shell model, as well as the value reported by Yasuda \etal~\cite{Yasuda:2010zz,Yasuda:2012}. This finding likely indicates that our fit assigns to the $1s_{1/2}$ state some strength belonging 
to other components of the SF. 

Overall, the results of previous studies listed in Table~\ref{table:Ar&Ca} suggest that argon and calcium are similar, although not all findings of our study  corroborate this suggestion. 
A firm assessment of the sources of the discrepancies is likely to require a more detailed analysis, based on an improved theoretical model, which is currently being 
developed.

%%%%%%%%%%%%%%%%%%%%%%%%%%%%%%%%%%%%%%%%%%%%%%%%%%%%%%%%%%%%%%
\section{Summary and Conclusions}\label{sec:Summary}
%%%%%%%%%%%%%%%%%%%%%%%%%%%%%%%%%%%%%%%%%%%%%%%%%%%%%%%%%%

The \isotope[40][18]{Ar}$(e,e^\prime p)$ data collected  
by experiment E12-14-012 at Jefferson Lab have been analysed to obtain the target spectral function,  describing the energy and momentum distribution of protons bound in the argon ground state. 

The model dependence involved in the determination of the spectral function\textemdash based on factorisation of the measured cross section\textemdash arises mainly from the treatment of FSI. 
The uncertainty associated with the choice of the proton optical potential, discussed in a previous article~\cite{Gu:2020rcp}, is included in the overall systematic uncertainty.

The results of our analysis provide important novel information, critical to the interpretation of events observed in liquid argon detectors. 

A quantitative understanding of the spectral function will allow to greatly improve the accuracy of neutrino energy reconstruction in long-baseline searches of neutrino 
 oscillations. In addition, it should be kept in mind that, being an intrinsic property of the target nucleus, the spectral function is relevant to the description of all reaction channels, 
 including quasielastic scattering, resonance production, and deep-inelastic scattering~\cite{Vagnoni:2017hll}.

The reduced differential cross sections\textemdash obtained from the data dividing out the elementary electron-nucleon 
cross section and a trivial kinematic factor\textemdash turn out to  be a function of missing energy and missing momentum only, that has been fitted using a model 
spectral function. The effects of FSI, which are known to be significant in $(e,e^\prime p)$ reactions, have been taken into account within the well established framework 
based on DWIA.  

The comparison between data and results of the MC simulations has been performed in a broad range of missing energies, extending from proton knockout threshold
to $E_m \sim 80$ MeV. The emerging picture, showing a good overall agreement, supports the validity of the theoretical basis of our analysis.    

We have been able to determine position and width of the peaks corresponding to  
shell model states, and to estimate the corresponding spectroscopic strengths. The comparison 
between the results of our analysis and those of 
past experiments aimed at extracting occupation probabilities in argon and calcium
is very encouraging, although some unresolved issues remain to be addressed. 

A more accurate determination of the argon spectral function\textemdash allowing a clearcut identification of  
contributions associated with both single-particle states and the correlation continuum\textemdash will require a more advanced theoretical model of the 
energy and momentum distributions, as well as a refined implementation of the DWIA, in which the
large overlaps between the momentum distributions of different shell model states are properly accounted for.

The extraction of the spectral function reported in this article\textemdash providing a satisfactory description of the proton energy and momentum distribution needed for the description of (anti)neutrino interactions in argon\textemdash should be seen as the achievement of first goal of the experiment E12-14-012. However, the cross section of the process $\isotope[48][22]{Ti}(e,e^\prime p)$ has also been measured. Owing to the correspondence between the proton spectrum of titanium and the neutron spectrum of argon, the determination of the proton spectral function of titanium will contribute the complementary information needed to describe neutrino interactions. The analysis of titanium data
is under way, and will be discussed in a forthcoming publication.

%%%%%%%%%%%%%%%%%%%%%%%%%%%%%%%%%%%%%%%%%%%%%%%%%%%%%%%%%%%%%%%%%%%%%%%%%%%%%%%%%%%%%%%%%%%%%%%%%%%
\vspace{1cm}
\begin{acknowledgments}
\par We acknowledge the outstanding support from the Jefferson Lab Hall A technical staff, target group and Accelerator Division. This experiment was made possible by Virginia Tech, the National Science Foundation under CAREER grant No. PHY$-$1352106 and grant No. PHY$-$1757087. This work was also supported by the DOE Office of Science, Office of Nuclear Physics, contract DE-AC05-06OR23177, under which Jefferson Science Associates, LLC operates JLab, DOE contract DE-FG02-96ER40950, DE-AC02-76SF00515, DE-SC0013615 and by the DOE Office of High Energy Physics, contract DE-SC0020262.
\end{acknowledgments}

%
%
%
%\newpage
%
%-----------------------------------------------------------------------------------------------------------------------------------------------------------------------%

\end{document}